\documentclass[twocolumn]{aastex631}

\usepackage[T5,T1]{fontenc}
\usepackage{savesym}
\usepackage{amsmath}
\usepackage{enumitem}
\usepackage{booktabs}
\usepackage{multirow}
\usepackage{threeparttable}
\usepackage[printonlyused,withpage]{acronym}

\newcommand\ts{\mathcal{TS}}

\shortauthors{Abbasi et al.}

\begin{document}

\title{Constraints on populations of neutrino sources from searches in the directions of IceCube neutrino alerts}
\email{analysis@icecube.wisc.edu}

\affiliation{III. Physikalisches Institut, RWTH Aachen University, D-52056 Aachen, Germany}
\affiliation{Department of Physics, University of Adelaide, Adelaide, 5005, Australia}
\affiliation{Dept. of Physics and Astronomy, University of Alaska Anchorage, 3211 Providence Dr., Anchorage, AK 99508, USA}
\affiliation{Dept. of Physics, University of Texas at Arlington, 502 Yates St., Science Hall Rm 108, Box 19059, Arlington, TX 76019, USA}
\affiliation{CTSPS, Clark-Atlanta University, Atlanta, GA 30314, USA}
\affiliation{School of Physics and Center for Relativistic Astrophysics, Georgia Institute of Technology, Atlanta, GA 30332, USA}
\affiliation{Dept. of Physics, Southern University, Baton Rouge, LA 70813, USA}
\affiliation{Dept. of Physics, University of California, Berkeley, CA 94720, USA}
\affiliation{Lawrence Berkeley National Laboratory, Berkeley, CA 94720, USA}
\affiliation{Institut f{\"u}r Physik, Humboldt-Universit{\"a}t zu Berlin, D-12489 Berlin, Germany}
\affiliation{Fakult{\"a}t f{\"u}r Physik {\&} Astronomie, Ruhr-Universit{\"a}t Bochum, D-44780 Bochum, Germany}
\affiliation{Universit{\'e} Libre de Bruxelles, Science Faculty CP230, B-1050 Brussels, Belgium}
\affiliation{Vrije Universiteit Brussel (VUB), Dienst ELEM, B-1050 Brussels, Belgium}
\affiliation{Department of Physics and Laboratory for Particle Physics and Cosmology, Harvard University, Cambridge, MA 02138, USA}
\affiliation{Dept. of Physics, Massachusetts Institute of Technology, Cambridge, MA 02139, USA}
\affiliation{Dept. of Physics and The International Center for Hadron Astrophysics, Chiba University, Chiba 263-8522, Japan}
\affiliation{Department of Physics, Loyola University Chicago, Chicago, IL 60660, USA}
\affiliation{Dept. of Physics and Astronomy, University of Canterbury, Private Bag 4800, Christchurch, New Zealand}
\affiliation{Dept. of Physics, University of Maryland, College Park, MD 20742, USA}
\affiliation{Dept. of Astronomy, Ohio State University, Columbus, OH 43210, USA}
\affiliation{Dept. of Physics and Center for Cosmology and Astro-Particle Physics, Ohio State University, Columbus, OH 43210, USA}
\affiliation{Niels Bohr Institute, University of Copenhagen, DK-2100 Copenhagen, Denmark}
\affiliation{Dept. of Physics, TU Dortmund University, D-44221 Dortmund, Germany}
\affiliation{Dept. of Physics and Astronomy, Michigan State University, East Lansing, MI 48824, USA}
\affiliation{Dept. of Physics, University of Alberta, Edmonton, Alberta, Canada T6G 2E1}
\affiliation{Erlangen Centre for Astroparticle Physics, Friedrich-Alexander-Universit{\"a}t Erlangen-N{\"u}rnberg, D-91058 Erlangen, Germany}
\affiliation{Physik-department, Technische Universit{\"a}t M{\"u}nchen, D-85748 Garching, Germany}
\affiliation{D{\'e}partement de physique nucl{\'e}aire et corpusculaire, Universit{\'e} de Gen{\`e}ve, CH-1211 Gen{\`e}ve, Switzerland}
\affiliation{Dept. of Physics and Astronomy, University of Gent, B-9000 Gent, Belgium}
\affiliation{Dept. of Physics and Astronomy, University of California, Irvine, CA 92697, USA}
\affiliation{Karlsruhe Institute of Technology, Institute for Astroparticle Physics, D-76021 Karlsruhe, Germany }
\affiliation{Karlsruhe Institute of Technology, Institute of Experimental Particle Physics, D-76021 Karlsruhe, Germany }
\affiliation{Dept. of Physics, Engineering Physics, and Astronomy, Queen's University, Kingston, ON K7L 3N6, Canada}
\affiliation{Department of Physics {\&} Astronomy, University of Nevada, Las Vegas, NV, 89154, USA}
\affiliation{Nevada Center for Astrophysics, University of Nevada, Las Vegas, NV 89154, USA}
\affiliation{Dept. of Physics and Astronomy, University of Kansas, Lawrence, KS 66045, USA}
\affiliation{Department of Physics and Astronomy, UCLA, Los Angeles, CA 90095, USA}
\affiliation{Centre for Cosmology, Particle Physics and Phenomenology - CP3, Universit{\'e} catholique de Louvain, Louvain-la-Neuve, Belgium}
\affiliation{Department of Physics, Mercer University, Macon, GA 31207-0001, USA}
\affiliation{Dept. of Astronomy, University of Wisconsin{\textendash}Madison, Madison, WI 53706, USA}
\affiliation{Dept. of Physics and Wisconsin IceCube Particle Astrophysics Center, University of Wisconsin{\textendash}Madison, Madison, WI 53706, USA}
\affiliation{Institute of Physics, University of Mainz, Staudinger Weg 7, D-55099 Mainz, Germany}
\affiliation{Department of Physics, Marquette University, Milwaukee, WI, 53201, USA}
\affiliation{Institut f{\"u}r Kernphysik, Westf{\"a}lische Wilhelms-Universit{\"a}t M{\"u}nster, D-48149 M{\"u}nster, Germany}
\affiliation{Bartol Research Institute and Dept. of Physics and Astronomy, University of Delaware, Newark, DE 19716, USA}
\affiliation{Dept. of Physics, Yale University, New Haven, CT 06520, USA}
\affiliation{Columbia Astrophysics and Nevis Laboratories, Columbia University, New York, NY 10027, USA}
\affiliation{Dept. of Physics, University of Oxford, Parks Road, Oxford OX1 3PU, UK}
\affiliation{Dipartimento di Fisica e Astronomia Galileo Galilei, Universit{\`a} Degli Studi di Padova, 35122 Padova PD, Italy}
\affiliation{Dept. of Physics, Drexel University, 3141 Chestnut Street, Philadelphia, PA 19104, USA}
\affiliation{Physics Department, South Dakota School of Mines and Technology, Rapid City, SD 57701, USA}
\affiliation{Dept. of Physics, University of Wisconsin, River Falls, WI 54022, USA}
\affiliation{Dept. of Physics and Astronomy, University of Rochester, Rochester, NY 14627, USA}
\affiliation{Department of Physics and Astronomy, University of Utah, Salt Lake City, UT 84112, USA}
\affiliation{Oskar Klein Centre and Dept. of Physics, Stockholm University, SE-10691 Stockholm, Sweden}
\affiliation{Dept. of Physics and Astronomy, Stony Brook University, Stony Brook, NY 11794-3800, USA}
\affiliation{Dept. of Physics, Sungkyunkwan University, Suwon 16419, Korea}
\affiliation{Institute of Basic Science, Sungkyunkwan University, Suwon 16419, Korea}
\affiliation{Institute of Physics, Academia Sinica, Taipei, 11529, Taiwan}
\affiliation{Dept. of Physics and Astronomy, University of Alabama, Tuscaloosa, AL 35487, USA}
\affiliation{Dept. of Astronomy and Astrophysics, Pennsylvania State University, University Park, PA 16802, USA}
\affiliation{Dept. of Physics, Pennsylvania State University, University Park, PA 16802, USA}
\affiliation{Dept. of Physics and Astronomy, Uppsala University, Box 516, S-75120 Uppsala, Sweden}
\affiliation{Dept. of Physics, University of Wuppertal, D-42119 Wuppertal, Germany}
\affiliation{DESY, D-15738 Zeuthen, Germany}

\author[0000-0001-6141-4205]{R. Abbasi}
\affiliation{Department of Physics, Loyola University Chicago, Chicago, IL 60660, USA}

\author[0000-0001-8952-588X]{M. Ackermann}
\affiliation{DESY, D-15738 Zeuthen, Germany}

\author{J. Adams}
\affiliation{Dept. of Physics and Astronomy, University of Canterbury, Private Bag 4800, Christchurch, New Zealand}

\author{N. Aggarwal}
\affiliation{Dept. of Physics, University of Alberta, Edmonton, Alberta, Canada T6G 2E1}

\author[0000-0003-2252-9514]{J. A. Aguilar}
\affiliation{Universit{\'e} Libre de Bruxelles, Science Faculty CP230, B-1050 Brussels, Belgium}

\author[0000-0003-0709-5631]{M. Ahlers}
\affiliation{Niels Bohr Institute, University of Copenhagen, DK-2100 Copenhagen, Denmark}

\author[0000-0002-9534-9189]{J.M. Alameddine}
\affiliation{Dept. of Physics, TU Dortmund University, D-44221 Dortmund, Germany}

\author{A. A. Alves Jr.}
\affiliation{Karlsruhe Institute of Technology, Institute for Astroparticle Physics, D-76021 Karlsruhe, Germany }

\author{N. M. Amin}
\affiliation{Bartol Research Institute and Dept. of Physics and Astronomy, University of Delaware, Newark, DE 19716, USA}

\author{K. Andeen}
\affiliation{Department of Physics, Marquette University, Milwaukee, WI, 53201, USA}

% ...

\date{\today}
\collaboration{385}{IceCube Collaboration}
\noaffiliation

\begin{abstract}
Beginning in 2016, the IceCube Neutrino Observatory has sent out alerts in real time containing the information of high-energy ($E \gtrsim 100$~TeV) neutrino candidate events with moderate-to-high ($\gtrsim 30$\%) probability of astrophysical origin. In this work, we use a recent catalog of such alert events, which, in addition to events announced in real-time, includes events that were identified retroactively, and covers the time period of 2011-2020. We also search for additional, lower-energy, neutrinos from the arrival directions of these IceCube alerts. We show how performing such an analysis can constrain the contribution of rare populations of cosmic neutrino sources to the diffuse astrophysical neutrino flux. After searching for neutrino emission coincident with these alert events on various timescales, we find no significant evidence of either minute-scale or day-scale transient neutrino emission or of steady neutrino emission in the direction of these alert events. %Assuming sources have the same luminosity, an $E^{-2.5}$ neutrino spectrum and number densities that follow star-formation rates, this study shows that a population of neutrino sources has to be more numerous than $7\times 10^{-9}~\textrm{Mpc}^{-3}$, or $3\times 10^{-7}~\textrm{Mpc}^{-3}$ if number densities instead have no cosmic evolution, to account for the complete astrophysical neutrino flux.
This study also shows how numerous a population of neutrino sources has to be to account for the complete astrophysical neutrino flux. Assuming sources have the same luminosity, an $E^{-2.5}$ neutrino spectrum and number densities that follow star-formation rates, the population of sources has to be more numerous than $7\times 10^{-9}~\textrm{Mpc}^{-3}$. This number changes to $3\times 10^{-7}~\textrm{Mpc}^{-3}$ if number densities instead have no cosmic evolution.

\end{abstract}


\section{Introduction} \label{sec:intro}
Nearly a decade after the detection of a diffuse flux of astrophysical neutrinos~\citep{IceCube:2013low}, the origins of this flux largely remains a mystery. However, to that end, IceCube -- a cubic kilometer neutrino telescope operating at the geographic South Pole -- found compelling evidence that a blazar, TXS~0506+056, is a source of high-energy neutrinos~\citep{IceCube:2018cha,IceCube:2018dnn}, 
% that two objects, the blazar TXS0506+056 and the nearby Seyfert-II galaxy NGC 1068, are sources of high-energy neutrinos
though this object alone can only explain a small portion of the diffuse flux. Pinpointing more sources of cosmic neutrinos, or understanding the populations of sources which contribute to the overall measured diffuse flux~\citep{IceCube:2015gsk,IceCube:2020acn,IceCube:2020wum,IceCube:2021uhz} could prove pivotal in  understanding the processes behind the acceleration and propagation of high-energy cosmic rays.

The identification of the blazar TXS~0506+056, as well as some more recent claims of possible neutrino sources, e.g.~\cite{Stein:2020xhk,Franckowiak:2020qrq}, were enabled in part because of the correlations of these sources with public IceCube neutrino alerts~\citep{IceCube:2016cqr}. In addition to searching for correlations with individual alerts, some have looked for correlations between catalogs of sources and neutrino alerts, e.g.~\cite{Plavin:2020emb}. These high-energy neutrino alerts are often used to trigger follow-up observations because of their significant probabilities of astrophysical origin. By only looking at neutrino candidate events with high estimated initial energies ($E_{\nu} \gtrsim 100$~TeV), the typically overwhelming backgrounds from atmospheric cosmic-ray interactions can be suppressed. 

\begin{figure*}[t]
    \centering
    \includegraphics[width=0.7\textwidth]{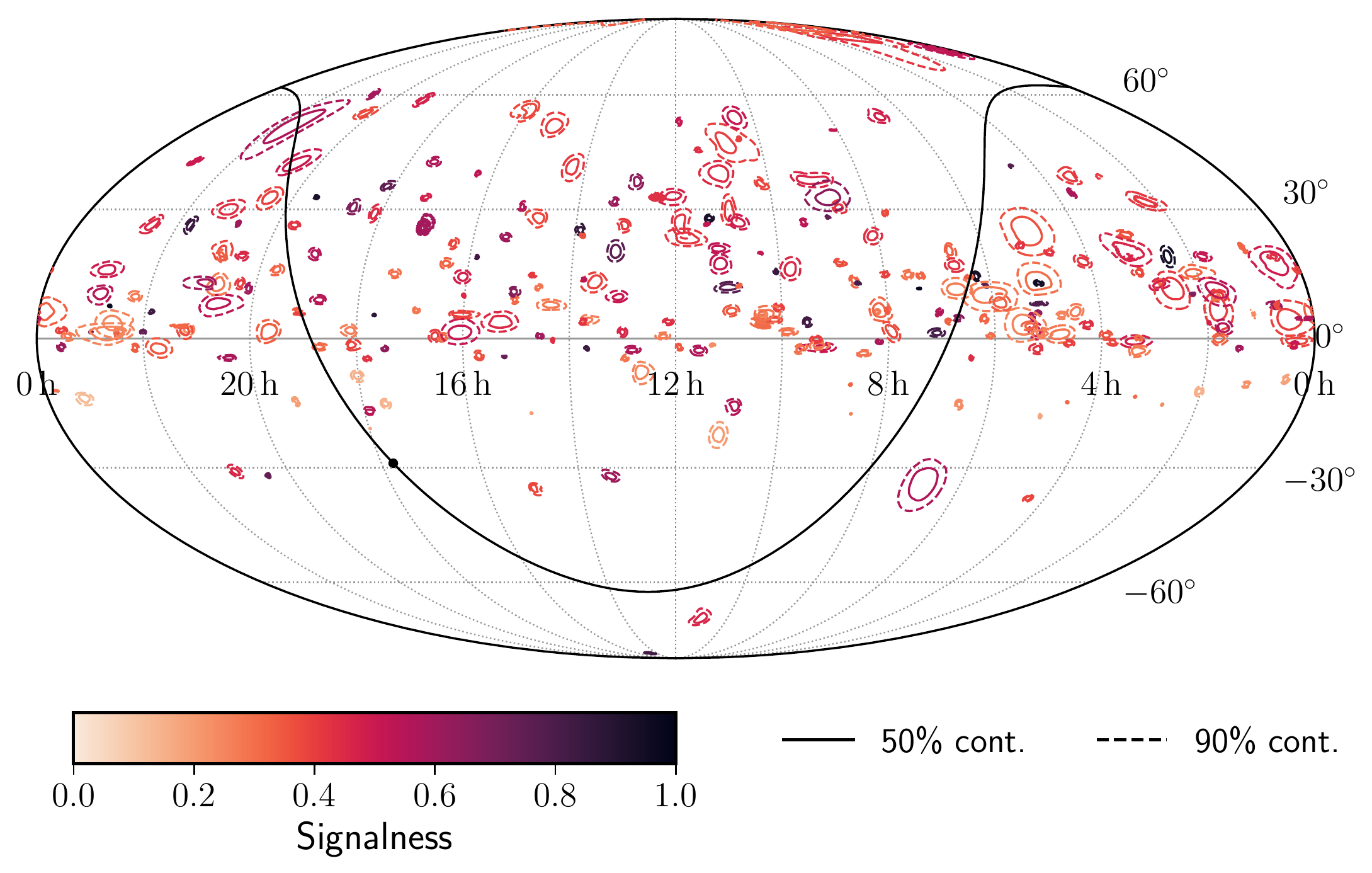}
    \caption{Skymap in Equatorial coordinates (J2000) of all neutrino candidate alerts used in the analysis described in Section~\ref{sec:analysis}. Contours denote the 50\% (solid) and 90\% (dashed) containment based on
    %off of 
    rescalings of the likelihood space according to resimulations of the event IceCube-160427A~\citep[][ see Section ~\ref{sec:v2_alerts} for more details]{Pan-STARRS:2019szg}. The color indicates the signalness of each alert event, described in the text. The Galactic Plane and Galactic Center are shown as a black solid line and dot, respectively.}
    \label{fig:all_sky_contours}
\end{figure*}

While it is clear that using high-energy neutrino alerts to trigger multi-wavelength (MWL) observations is a promising and fruitful way to identify cosmic neutrino sources, it is not without its limitations. First, it is not clear which astrophysical objects are sources of high-energy neutrinos. This, in combination with the non-negligible localization uncertainties for neutrino events, can lead to source confusion when using pointed MWL observations, especially as recent limits suggest that any neutrino source population responsible for a significant fraction of the diffuse flux must be fairly numerous~\citep{IceCube:2018ndw}. Second, there is not a consensus on the types of MWL emission that one expects to see with high-energy neutrinos. 
%For example, while the gamma-ray flare of the object TXS~0506+056 coincident with IC170922A in 2017 lent evidence for a connection~\citep{IceCube:2018dnn}, other works that have studied a potential correlation between neutrino emission and gamma-ray emission in blazars have yielded constraining upper limits, e.g.~\cite{Murase:2016gly,IceCube:2016qvd,Murase:2018iyl,Yuan:2019ucv,Oikonomou:2019djc}, and more recent studies have suggested correlating with time-dependent behavior at other frequencies might be more telling for correlating with neutrino emission~\citep{Kun:2020njy}. 
For example, works motivated by the IC170922A event in 2017 found to be coincident with a gamma-ray flare of TXS~0506+056 providing evidence for a connection ~\citep{IceCube:2018dnn}, have studied for potential correlation between neutrino emission and gamma-ray emission in blazars. These works have yielded constraining upper-limits, e.g.~\cite{Murase:2016gly,IceCube:2016qvd,Murase:2018iyl,Yuan:2019ucv,Oikonomou:2019djc}, with more recent studies suggesting that correlation with time-dependent behavior at other frequencies might be more telling for correlating with neutrino emission~\citep{Kun:2020njy}. 
Third, trying to consistently model neutrino emission from sources based on alert and MWL observations are subject to the Eddington bias~\citep{Strotjohann:2018ufz}, because alert events are likely from sources where, although the joint contribution from a population of sources to the diffuse flux might be large, alert events from individual sources are likely Poisson overfluctuations. Lastly, pointed MWL follow-up observations for alert events can be expensive, and not all alerts are followed up by instruments either because of low event-by-event astrophysical probabilities or because of observational constraints.

In this work, we rely on a different method of using neutrino alerts to search for astrophysical neutrino sources -- by following up neutrino candidate alert events using lower-threshold and higher statistics neutrino data. We use the word "followup" to denote performing analyses that use other neutrino data to search for astrophysical signals in the vicinity of the neutrino candidate alert. In addition to circumventing many of the difficulties that surround using MWL observations to followup neutrino alerts, this search strategy complements the all-sky searches and catalog based searches by reducing 
%also reduces 
the number of unique locations on the sky which need to be investigated, without requiring a fixed hypothesis on the astrophysical source class. This large trials-factor from the ``look-elsewhere effect'' typically degrades the sensitivity of 
%neutrino source 
searches that look for possible neutrino sources at every location on the sky. This could be one of the reasons why no source has been detected at above the $3\sigma$ level from these types of searches~\citep{IceCube:2019cia}. 

In this paper, we report the results of searches for neutrino sources in the directions of IceCube neutrino candidate alerts. In Section~\ref{sec:v2_alerts}, we describe the data samples used and then we outline the analysis techniques to analyze these data in Section~\ref{sec:analysis}. In Section~\ref{sec:population}, we show how we combine the results from the individual analyses to search for an overall excess of lower-energy neutrinos in the direction of high-energy neutrino alerts. After discussing the results of the analysis in Section~\ref{sec:results}, we show how these results can be used to constrain populations of neutrino sources in Section~\ref{sec:constraints}. 

\section{Data Samples} 
\label{sec:v2_alerts}
The IceCube Neutrino Observatory is a gigaton-scale Cherenkov detector embedded in the ice at the geographic South Pole~\citep{IceCube:2016zyt}. The detector consists of 5,160 Digital Optical Modules (DOMs) dispersed on 86 ``strings'' arrayed in a hexagonal grid and deployed at depths of 1450-2450~m beneath the ice surface. Each DOM contains a 10~inch photomultiplier tube suited to detect optical Cherenkov photons~\citep{IceCube:2010dpc} as well as read-out and digitization electronics~\citep{IceCube:2008qbc}. Neutrinos are detected indirectly via the Cherenkov radiation produced from relativistic charged particles created by deep inelastic neutrino nucleon interactions in the surrounding ice or nearby bedrock beneath IceCube. 

Although sensitive to all flavors of neutrino interactions, this study relies only on muon track events from muon-neutrino charged current interactions as well as a 10\% contribution from muonic tau decays from charged current tau-neutrino interactions. These ``track'' events enable a better angular resolution than the other event type, ``cascades'' (from charged current electron- and tau-neutrino interactions or neutral current interactions of all flavors), at the cost of a poorer energy resolution. The angular resolution of track events is preferable when searching for point sources in the region of the sky where most of the alert events are detected.

In addition to neutrinos from astrophysical sources, IceCube detects many neutrinos and muons from cosmic-ray interactions in the atmosphere. In the southern celestial hemisphere, the events detected by IceCube are dominated by atmospheric muons, with events from atmospheric neutrinos still occurring at rates a few orders of magnitude larger than astrophysical neutrinos. In the northern celestial hemisphere, the atmospheric muons are attenuated by the Earth, and the rate is dominated by atmospheric neutrinos. 

The analysis presented here leverages the strengths of two different IceCube data streams: the alert-event stream and the ``gamma-ray follow-up'' (GFU) stream \citep{gfu_2016}. Both of these event selections try to isolate neutrino candidate events with low latency, enabled by a realtime alert infrastructure that began sending alerts publicly in April 2016, and are described in full in~\cite{IceCube:2016cqr}. However, the selection criteria used to identify alert events was revisited in 2019~\citep{Blaufuss:2019fgv} to expand the alert program, and the alert stream now consists of two unique channels: ``Gold'' events which have an average astrophysical signal purity above 50\%, and ``Bronze'' which have an average astrophysical signal purity above 30\%. These event-by-event astrophysical purities are calculated by finding the event ``signalness'', $\mathcal{S}$, which is the ratio of the expected number of events from signal to the expected total number of events (signal plus background) at a given declination with energies greater than the reconstructed energy of the event~\citep{Blaufuss:2019fgv}. The final rate is approximately 10 events per-year in the ``Gold'' selection, and 30 events overall in the ``Gold'' and ``Bronze'' selection. Signalness is dependent on the assumed spectral index. To avoid this dependence, the alert stream effective area, which is a function of the energy, is used for the analysis while treating all events to be on the same footing.

Whereas the alert stream is optimized for finding individual events with moderate-to-high probability of astrophysical origin, the GFU sample is focused on optimizing sensitivity to short timescale transients. When searching for transient neutrino emission, the effective background of the analysis is reduced because each analysis only looks at a narrow swath of sky and for a limited period of time. Because of this, the cuts for the event selection can be looser than for the alert stream. The final all-sky rate is $\sim 6.7$~mHz (approximately $2\times 10^{5}$ events per year). While the vast majority of this is from atmospheric backgrounds, the effective area for astrophysical neutrinos is significantly larger with the GFU sample than with the alert sample. For example, consider the quantity
\begin{equation}
\begin{aligned}
\label{eq:nevents}
    \langle \mathcal{N}^{\mathrm{stream}}(\delta, \gamma, \phi_0) \rangle = & \int_0^{\infty} \phi_0 \bigg(  \frac{E}{1\;\mathrm{TeV}}\bigg)^{-\gamma} \\ & \times A_{\mathrm{stream}}^{\mathrm{eff}}(\delta, E) dE ,
\end{aligned}
\end{equation}
which is the expected number of detected events for a source at a declination $\delta$ with an $E^{-\gamma}$ spectrum with normalization $\phi_0$ in a given event stream (either the alert stream or GFU). The term $A_{\mathrm{stream}}^{\mathrm{eff}}$ describes the effective area of the event selection. Then, for a source with $\gamma=2.5$ ($\gamma=2.0$), the ratio of the expected number of events in the GFU stream to the alert stream is 97 (13) for a source at $\delta=0^{\circ}$ and 270 (40) for a source in the northern sky. 
%This means that if a source is bright enough for there to be an expectation of observing one alert event, then using the GFU event selection to search for a source in the same direction of the alert could have tens to hundreds of lower-energy events coincident with the source.
This means that if a source, in the same direction of the alert, is bright enough for there to be an expectation of observing one alert event, then a search with the GFU event selection could result in tens to hundreds of lower-energy events coincident with the source.
This also helps to remove some of the aforementioned Eddington bias, as using a larger statistics sample can provide a better estimate of the true source flux. It is worth noting that most (88\%) of the alert events are included in the GFU sample. When we perform a followup for each alert, we remove the alert that prompted the analysis from the followup.

In addition to events detected in real time after the creation of the alert selection criteria, archival IceCube data dating back to 2011 have been processed with the same sets of cuts used for the alert and GFU streams. Overall, our datasets span May 13, 2011 to December 31, 2020, and the final alert stream has 275 events and the GFU selection has $1.8\times 10^6$ events from this period.

Complementing the reconstructions that were applied to each event in the GFU stream, described in~\cite{IceCube:2016cqr}, alert events have an additional reconstruction applied to them. This includes effects from systematic uncertainties after they have been initially circulated to the public.
%Complementing the reconstructions that were applied to each event in the GFU stream, which are described in~\cite{IceCube:2016cqr}, alert events have an additional reconstruction applied to them which includes effects from systematic uncertainties after they have been initially circulated to the public.
Because of the computational constraints of this reconstruction, it is reserved for only the events passing alert cuts. 
In order to calculate uncertainty contours for the directional reconstruction of each alert event, they are compared against re-simulations of other high-energy track events which varied the allowed models of the optical properties of the deep glacial ice. This process is described in full in~\cite{IceCube:2021uwf}, and it allows one to quote 50\% and 90\% containment contours for the directional reconstruction of the event. 
%In order to calculate uncertainty contours for the directional reconstruction of each alert event, they are compared against re-simulations of other high-energy track events which varied the allowed models of the optical properties of the deep glacial ice, this process is described in full in~\cite{IceCube:2021uwf}, and it allows one to quote 50\% and 90\% containment contours for the directional reconstruction of the event.
This process was first applied for the event IceCube-160427A, for which a likely unrelated coincident supernova was found by Pan-STARRS~\citep{Pan-STARRS:2019szg}. For events detected in real time, bounds to these these contours are typically circulated to the community within a few hours of the initial detection of the event. 

Figure~\ref{fig:all_sky_contours} shows the sample of alert-quality events that we analyze with our followup analyses by showing the 50\% and 90\% contours calculated using the procedure outlined above. A few events from the full catalog of alert events are excluded from our analysis because the computation time for each followup scales with the alert uncertainty size, and some alert events had exceptionally large angular uncertainties which prevented us from performing enough pseudo-experiments to characterize our background expectations. Although those events are excluded from the Figure~\ref{fig:all_sky_contours}, we tabulate them in the full table of results in Appendix~\ref{app:results_table}. Additionally, when searching for emission on short timescales, some alerts are excluded just from the transient analyses we perform if there was a significant period of detector downtime near the time of the alert. These are also mentioned in the Table in Appendix~\ref{app:results_table}.

\section{Analysis Method} \label{sec:analysis}

The analyses we perform rely on an unbinned maximum likelihood technique that is well-established in many searches~\citep{Braun:2008bg}. For recent examples of such analyses, see e.g.~\cite{IceCube:2019cia,IceCube:2018ndw}. For each of these analyses, we are searching for GFU events that are spatially clustered and coincident with alert events. This is done by checking if the alert event which prompted the followup is also in the GFU sample, and if it is, we remove it from the search. As we do not know if populations of neutrino sources that are responsible for the diffuse flux are transient sources or sources that emit over long timescales, we perform several analyses that test different temporal hypotheses. Specifically, we perform likelihood analyses on three different timescales: (1) centered on the alert time and searching for spatially coincident events in a time-range of $\pm 500$~s, (2) centered on the alert time and searching for spatially coincident events in a time-range of $\pm 1$~day, and (3) searching for spatially coincident events during the entire livetime of the GFU sample. The durations of the transient analyses were chosen to strike a balance between theoretical and empirical constraints. The shorter timescale ($\pm 500$~s) has been suggested as a conservative approximate timescale for neutrino emission from compact binary mergers~\citep{Baret:2011tk,IceCube:2020xks}, and has been used in many IceCube searches for transients. The 2~day timescale reflects the longest timescales of neutrino emission proposed for neutrino emission from some short timescale transients, e.g. Fast Radio Bursts~\citep{Metzger:2020byf}, and it is also the maximal time window in which our analysis remains sensitive to individual coincident events. It is worth noting that searches that look for GFU events that are clustered in time but are not necessarily coincident in time with the alert event (so-called ``flare analyses'') are not performed in this work, although a dedicated analysis repeating the flare analysis that identified the 2014-2015 flare from TXS~0506+056 are in development~\citep{IceCube:2021kxg}.

At the core of each analysis presented here is the same likelihood framework, with some differences which we outline below. Given a location on the sky with Equatorial coordinates, $\vec{\Omega} = (\alpha, \delta)$, the likelihood consists of the weighted sum of a signal probability distribution function (PDF), $S$, and a background PDF, $B$, and is given by
\begin{align}
\label{eq:likelihood}
    \mathcal{L}\left(\vec{\Omega}, n_{s}, \gamma\right)= \lambda \prod_{i=1}^{N} & \Bigg(\frac{n_{s}}{N} \cdot S\left(\vec{\Omega}_{i}, E_{i}, \sigma_{i} \mid \gamma, \vec{\Omega}\right) \\ & \quad + \left(1-\frac{n_{s}}{N}\right) \cdot B\left(\delta_{i}, E_{i}\right) \Bigg) \; . \nonumber
\end{align}
The index $i$ iterates over all $N$ neutrino candidate events in the GFU sample and $n_s$ is the number of signal neutrino events. The signal and background PDFs, $S$ and $B$, are functions of four observables associated with each event: the reconstructed right ascension and declination, $\vec{\Omega}_i = (\alpha_i, \delta_i)$, the reconstructed energy, $E_i$, and the angular uncertainty $\sigma_i$. Both $S$ and $B$ are themselves products of two terms: energy and spatial PDFs. The signal energy PDF is a declination-dependent reconstructed energy distribution, where the underlying neutrino flux is modeled as a power-law,
\begin{equation}
    \frac{d N}{dE dA dt} = \phi_0 \times \bigg( \frac{E}{1\mathrm{TeV}}\bigg)^{-\gamma} \; ,
\end{equation}
where $\phi_0$ is the flux normalization and $\gamma$ is the spectral index. The spatial term of the signal PDF is modeled as a Gaussian with width $\sigma_i$, given by the angular uncertainty estimator of each neutrino candidate event in the GFU sample. The energy and spatial components of the background PDF are determined as functions of the reconstructed energy, $E_i$ and declination, $\delta_i$, for each event. A more thorough description of how the signal and background PDFs are calculated is provided in~\cite{IceCube:2016tpw}.

\begin{figure*}[t]
    \includegraphics[width=0.99\textwidth]{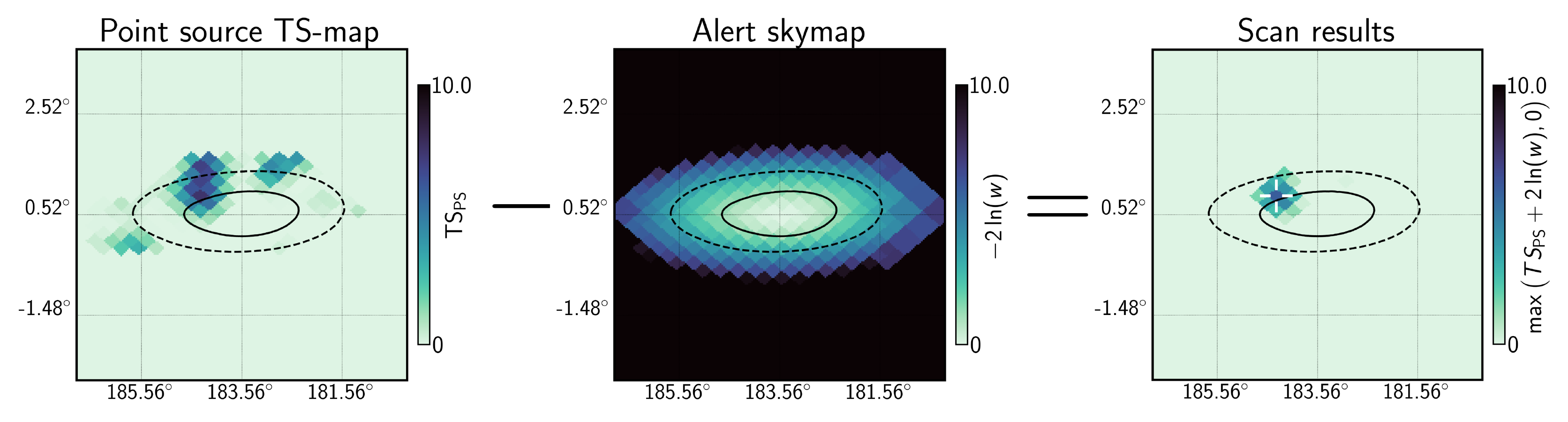}
    \caption{Schematic of the likelihood analysis described in Section~\ref{sec:analysis}. For each neutrino alert candidate event, the local point source $\ts$-map is calculated after excluding the alert event from the dataset (left). From this map we then subtract a term that is constructed from the likelihood map derived from a more sophisticated and dedicated reconstruction of the alert event (middle) to produce an overall map (right) from which the maximum is used as the analysis test statistic, $\ts$, corresponding to the location denoted here with a white cross hair. The alert shown here, Run120027:Event12133428, was chosen just for visualization purposes, and black contours show 50\% (solid) and 90\% (dashed) containment derived from the map displayed in the middle. This shows a time-integrated followup, but we repeat the procedure for the shorter timescale followups as well.}
    \label{fig:analysis_schematic}
\end{figure*}

The difference in the transient analyses and the time-integrated analysis is encapsulated in the $\lambda$ term in Eq.~\ref{eq:likelihood}. When searching for short timescale transient emission, the data is divided into the time period of interest, ``on-time'' data, and the remaining, ``off-time'', data. We then set $\lambda$ to a Poisson probability, $\lambda = (n_s + n_b)^N \exp(-n_s - n_b) / N!$, where $n_b$ is the expected number of background events on the entire sky in the time-window estimated using the surrounding off-time data. This ``extended likelihood'' methodology~\citep{1990NIMPA.297..496B} has been a feature of many analyses searching for short timescale neutrino emission~\citep{IceCube:2020mzw,IceCube:2017amx,IceCube:2019acm,IceCube:2020xks}. For the time-integrated analysis looking for steady emission over the entire livetime of the GFU sample, we set $\lambda = 1$.

The best-fit signal parameters at a given source position are then obtained using the maximum-likelihood method. For the transient analyses, we only maximize the likelihood with respect to $n_s$, and we keep $\gamma$ fixed to 2.5. This is because in transient analyses we are looking for few individual coincident events, and it is not feasible to fit a spectral index from the observation of a single event. For the time-integrated analysis, we maximize with respect to both $n_s$ and $\gamma$. Using a likelihood ratio test, we calculate the point-source test statistics $\ts_{\mathrm{ps}}$, as
\begin{equation}
    \ts_{\mathrm{ps}} (\vec{\Omega}) = 2 \ln \bigg( \frac{\mathcal{L}(\vec{\Omega}, \hat{n}_s, \hat{\gamma})}{\mathcal{L}(\vec{\Omega}, n_s = 0)} \bigg) \; ,
\end{equation}
where $\hat{n}_s$ is the best-fit $n_s$ and $\hat{\gamma}$ is the best-fit $\gamma$ (or fixed to 2.5 for the transient analyses). Here, the null hypothesis is defined as $n_s=0$, representing the case of no neutrino source in the direction $\vec{\Omega}$.

Thus far, we have described how to search for a source at a location $\vec{\Omega}$. However, we would like to search for sources in the direction of IceCube alert events, and these alert events are not perfectly localized. In order to do this, we first divide the sky into a grid using the HEALpix scheme~\citep{Gorski:2004by} and calculate $\ts_{\mathrm{ps}}$ on each point of this grid, using a resolution of approximately 0.2$^{\circ}$. 

We then use the likelihood scan of each alert event to create a skymap. \cite{IceCube:2021uwf} details how 50\% and 90\% containment contours are calculated based on finding critical likelihood values. However, directly normalizing the likelihood space as a function of location on the sky would neglect these critical values. In order to obtain a skymap that
%is reflective of 
reflects the uncertainties dictated by systematic resimulation studies, we apply an order-preserving transformation to re-calibrate the likelihood space such that when we normalize the likelihood as a function of location on the sky, approximately 90\% of the skymap lies within the quoted 90\% systematic contour~\citep[Appendix I]{IceCube:2020wum}.

We then use an algorithm which effectively fits for the position of a point-like source in the environment of the alert direction. This is done, once we have a skymap normalized as a PDF as a function of location on the sky for an alert event, $P_{j}(\vec{\Omega})$. We include the skymap as a spatial constraint by multiplying it to the neutrino likelihood function via $\mathcal{L}\rightarrow \mathcal{L}\cdot P_{j}(\vec{\Omega})$. Because $P_{j}(\vec{\Omega})$ is independent of the variables which are floated when maximizing the likelihood, $n_s, \gamma$, the constraint manifests as adding a logarithmic term to the point-source test statistic defined for each grid point. 
%Note, that this algorithm is effectively fitting for the position of a point-like source in the environment of the alert direction. 
Finally, the test statistic from each alert followup is given by
\begin{equation}
    \ts = \max_{\vec{\Omega}} \Bigg( \ts_{\mathrm{ps}}(\vec{\Omega}) + 2\ln(w(\vec{\Omega}))  \Bigg) \; ,
\end{equation}
where $w(\vec{\Omega}) = P_{j}(\vec{\Omega}) / \max_{\vec{\Omega}}( P_{j}(\vec{\Omega}))$. Normalizing the skymap term, $w$, based on the maximum value of the skymap is just a choice of convention, and different convention choices add a constant term to the test statistic which can be omitted when calculating significances and $p$-values based on pseudo-experiments which are unique to each individual alert followup (see below). 
The analysis process is shown schematically in Figure~\ref{fig:analysis_schematic}, which shows the method to calculate the maximum test statistic for each alert followup. This is done by calculating the local point-source test statistic map and adding to it a logarithmic penalty term from the skymap. 
%The analysis process is shown schematically in Figure~\ref{fig:analysis_schematic}, which shows how for each alert followup, we calculate the local point-source test statistic map and add to it a logarithmic penalty term from the skymap in order to find the maximum test statistic for each alert followup. 
This procedure was initially developed in the context of searching for joint sources of ultra-high-energy cosmic rays and neutrinos~\citep{ANTARES:2022pdr} and has been used to search for neutrinos coincident with ANITA events~\citep{IceCube:2020gbx}, gravitational waves~\citep{IceCube:2020xks}, and gamma-ray bursts~\citep{grb_2022_icecube}. We then repeat this followup procedure for every alert in our catalog and for each of the 3 timescales described above. 

In order to calculate significances for each of the alert followups, we compare each observed test statistic to pseudo-experiments with {\it scrambled} data. For all pseudo-experiments, the PDFs in the likelihood and the alert skymaps are kept fixed.
%are kept fixed and the alert skymaps are kept fixed. 
We then randomize the GFU data in right ascension. We also perform tests to see how well the analysis is able to recover simulated signal. We randomize the arrival directions in the same method as we do for background-only pseudo-experiments, but in these cases, we also inject signal events assuming a true source position that is sampled from the skymap of the alert event. We can summarize the analysis performance by calculating a ``sensitivity'' for each followup, defined as the expected median one-sided Neyman upper limit at 90\% confidence under the assumption of the null hypothesis (no source in the direction of the alert event). %In general, the sensitivities are comparable to searching for a point source with no localization uncertainty. 
As some of the alert events have extremely large localization uncertainties, this increases the effective background and degrades the sensitivity of the analysis. For the short timescale analyses, this results in a sensitivity that can be up to 10\% worse than in the case of no localization uncertainty, when assuming the source has an $E^{-2.5}$ spectrum. For the time-integrated case, the sensitivity (and correspondingly, the upper limits in the cases of non-detections) when searching for sources coincident with alert skymaps is anywhere from 10\% to a factor of 2 worse than the localized point source case. In the case of potential detections, the greater effective background as compared to the localized point source scenario can have a larger effect on the signal strength required to confidently detect a source. For example, for most of the alert skymaps, the signal strength required to detect a source at the 3$\sigma$ level, before trials correction, using this skymap construction is at least a factor of 2 higher than the signal required to detect a source in the perfectly localized point source scenario. 

\section{Population Analysis} \label{sec:population}

Thus far, we have described how to search for significant correlations of GFU events with individual alert events. We perform the analysis described above for each of the alerts in the catalog and for each of our three analysis timescales. However, we would like to search for signals from populations of sources which might manifest as multiple alert followups that are individually not significant, but when combined, the total flux is detectable.

In order to accomplish this, we begin by calculating $p$-values for each of the individual followups using the procedure outlined in Section~\ref{sec:analysis}. Then, for each time window, we sort the $p$-values into a list, $p_{t, 1}, p_{t, 2}, \ldots, p_{t, N}$, where the index $t$ identifies the time-window\footnote{$N$ is not exactly 275 for each of the time-windows because some alert events were excluded from the analyses. This is taken into account when performing pseudo-experiments, and those events which are excluded are listed in Table~\ref{tab:results}.}. Under the background hypothesis ($n_s = 0$), these $N$ $p$-values are expected to be uniformly distributed between 0 and 1. The probability that $k$ or more $p$-values are smaller than or equal to $p_{t,k}$ is thus given by the binomial probability:
\begin{equation}
    \alpha_k = \sum_{m=k}^{N} \frac{N !}{(N-m) ! m !} p_{k}^{m}\left(1-p_{k}\right)^{N-m} \; .
\end{equation}
We then evaluate this probability for all possible number of sources, $k$, to calculate the overall binomial $p$-value, $\alpha = \min_k \alpha_k$. In order to account for the fact that we performed multiple tests by finding the most significant number of sources, $k$, we repeat this entire procedure on ensembles of background pseudo-experiments. We can then calculate an overall analysis $p$-value for each time window by comparing $\alpha$ to the distribution of this value that we obtain from these pseudo-experiments. When performing these ensembles of background pseudo-experiments, we use the same sets of scrambled data for all of the alert followup analyses. This ensures that any correlations between overlapping alert contours is properly accounted for when calculating an overall analysis $p$-value.

In Section~\ref{sec:v2_alerts}, we discuss how each alert has a corresponding ``signalness'', $\mathcal{S}$. Preliminary versions of the analysis attempted to incorporate this signalness parameter into the overall population test statistic, so as to give more weight to
%more heavily weight 
alert events that have a higher probability of astrophysical origin. However, the signalness of each alert event is calculated under the assumption that the diffuse astrophysical neutrino flux is described by the single power law fit reported in~\cite{Haack:2017dxi}, and thus the signalness is sensitive to changes in the spectral shape of the diffuse astrophysical neutrino flux. For this reason, we instead treat all alert events on equal footing, regardless of signalness.

\section{Analysis Results}
\label{sec:results}

\begin{figure}
    \centering
    \includegraphics[width=0.46\textwidth]{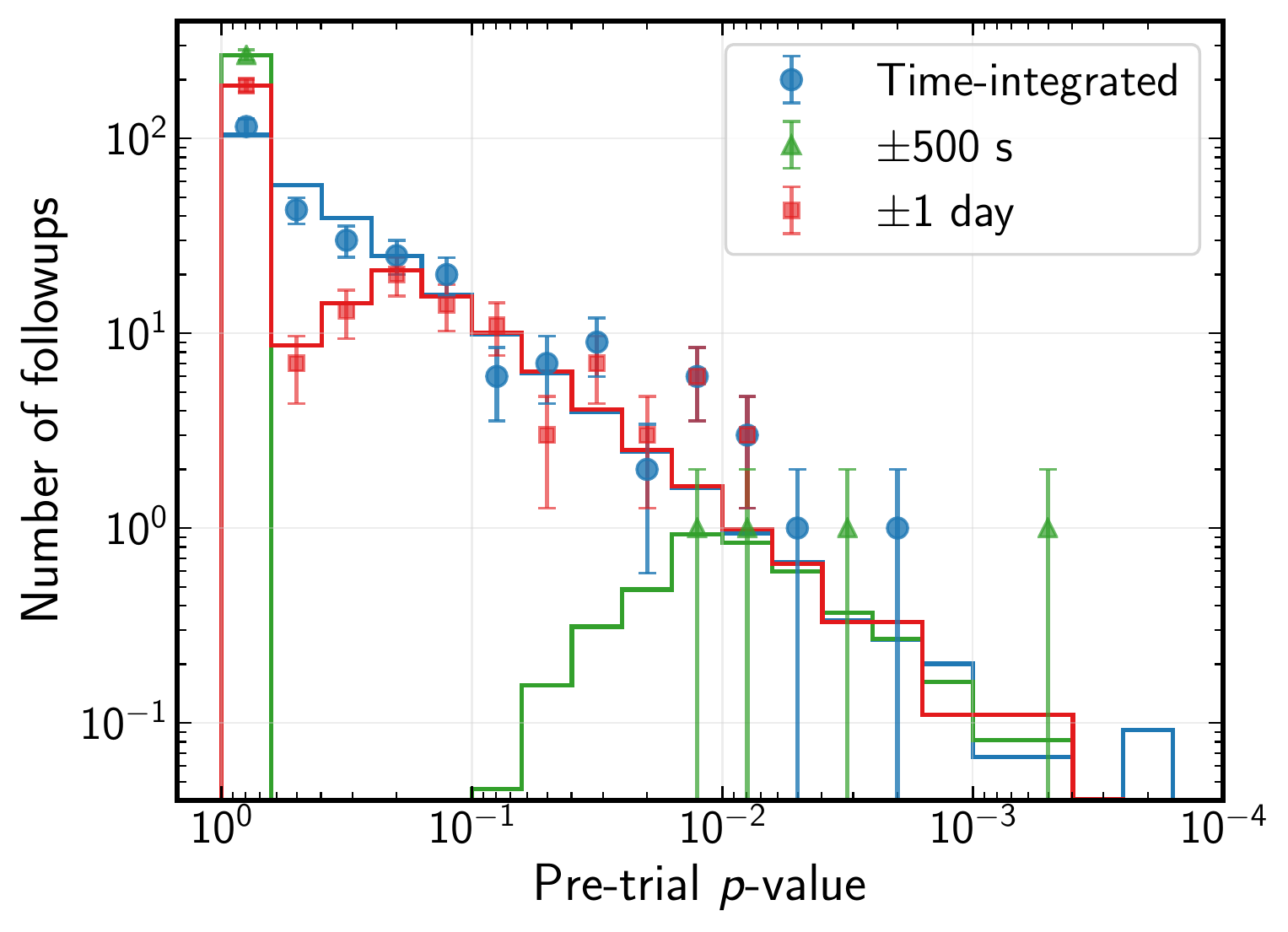}
    \caption{Results from all of the individual alert followup analyses. Observed $p$-values (markers) are compared to expectations from background-only pseudo-experiments (solid lines) for the time-integrated (blue), $\pm 500$~s (green), and $\pm 1$~day (red) analyses. $p$-values are shown before accounting for the trials factor accrued from the look elsewhere effect.}
    \label{fig:individual_results}
\end{figure}

The pre-trials $p$-values for all of the analyses are displayed in Figure~\ref{fig:individual_results}, and we also tabulate these values as well as the best-fit information from the time-integrated analysis in Table~\ref{tab:results} in Appendix~\ref{app:results_table}. We do not include the best-fit information for the short-timescale analyses in Table~\ref{tab:results}, as most of the best-fit number of events are $\hat{n}_s = 0$ and because we do not fit for the spectral index in those analyses. 

\begin{figure*}[t]
    \centering
    \includegraphics[width=0.44\textwidth]{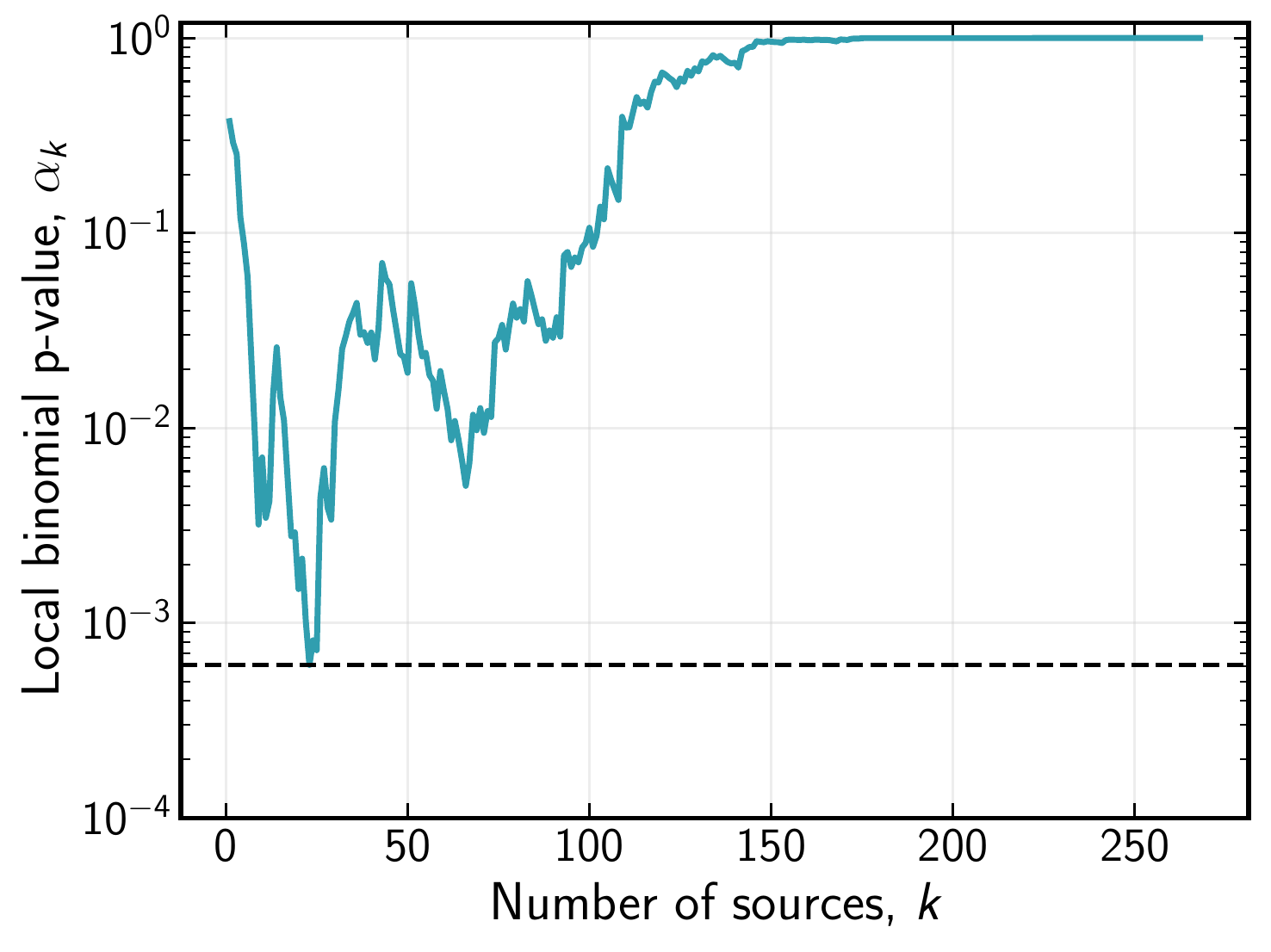}
    \includegraphics[trim={0 0 0 0.9cm},width=0.44\textwidth,clip]{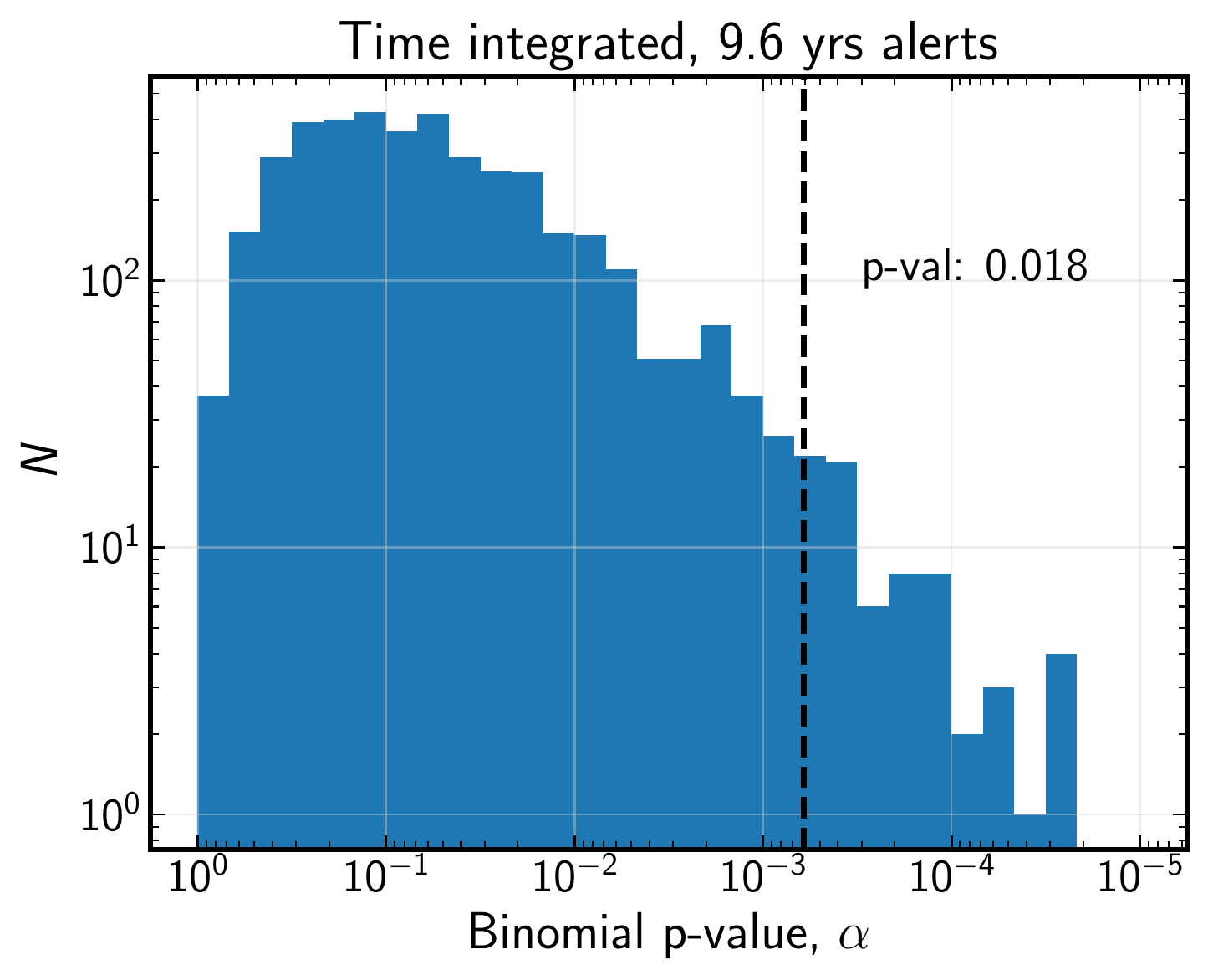}
    \caption{Binomial scan (left) and distribution of binomial p-values (right) calculated from pseudo-experiments using data scrambled in right ascension for the time-integrated analysis, as described in Section~\ref{sec:population}. In the binomial scan, the results for all alert followups is ordered by decreasing significance, and for each possible number of sources, $k$, we calculate the probability of obtaining $k$ results each with p-values less than or equal to $p_k$. The global minimum corresponds to $k=23$ and a binomial $p$-value of $\alpha=6.1\times 10^{-4}$ (black line). The observed binomial p-value from data is shown with a black line, compared to the expected background distribution in blue, and results in an overall p-value of 1.8\% for the time-integrated analysis.}
    \label{fig:binomial_distribution}
\end{figure*}

No individual alert followup is significant, especially after accounting for trials correction. The most significant followup comes from the $\pm 500$~s followup of the alert event 118342:24578488, which was detected on 2011-06-16. A single spatially coincident event with a reconstructed energy of $\sim$750~GeV that arrived 346~s after the alert event drives the significance, yielding $\hat{n}_s = 1$ and a pre-trials $p$-value of $6\times 10^{-4}$. After accounting for the fact that we performed nearly 275 analyses for each of the three time windows, this is consistent with background expectations, with a trials corrected $p$-value of approximately 40\%.

It is worth noting that the alert event that was identified as coincident with TXS~0506+056, IceCube-170922A, does not yield a significant result in this analysis (130033:50579430 in Table~\ref{tab:results} with pre-trial p value of 0.36 for the time-integrated case). However, this is completely consistent with what was reported in~\cite{IceCube:2018cha}, mainly because of the temporal hypothesis that is being tested. The analysis in~\cite{IceCube:2018cha} was searching for events that were clustered in time, and the clustering of events in 2014-2015 led to a more significant result than if the temporal hypothesis had been searching for time-integrated emission, as we do here. An analysis that performs the flare search that was done in~\cite{IceCube:2018cha} on the entire catalog of alert events is beyond the scope of this work, because of the computational constraints that come from the added dimensionality of maximizing the likelihood with respect to parameters that describe the temporal signal hypothesis. %~\citep{IceCube:2021kxg}. 
%Such an analysis is underway, and preliminary studies have already been reported~\citep{IceCube:2021kxg}. 
Additionally, that analysis was performed at the location of the source TXS~0506+056, whereas this analysis has increased background because we consider all locations within the uncertainty contour of the alert event. We do still fit $\hat{n}_s=17.5$ events, and the best-fit location is less than 0.2 degrees from the object TXS~0506+056. When comparing against other time-integrated analyses~\citep{IceCube:2019cia}, we fit a slightly softer spectrum ($\hat{\gamma}=2.75$), mainly because we exclude the alert event from our sample when performing the followup, and if it were included it would shift the spectral fit to be harder.

\begin{table*}
    \centering
    \begin{tabular}{ r | c c c }
    \hline
    \hline
        \textbf{Analysis time-window} & \textbf{Binomial p-value ($\alpha$)}  & \textbf{Number of sources ($k$)} & \textbf{Analysis p-value}  \\ \hline \hline
        $\pm 500$~s & $1.5\times 10^{-1}$ & 1 & 0.34 \\ \hline
        $\pm 1$~day & $1.0\times 10^{-2}$ & 19 & 0.12 \\ \hline 
        Time integrated & $6.1\times 10^{-4}$ & 23 & 0.018 \\ \hline \hline
    \end{tabular}
    \caption{Results from the binomial tests for each of the time-windows analyzed. We find all results to be consistent with background expectations.}
    \label{tab:binom_results}
\end{table*}

In addition to performing the individual followups, we can also perform the population tests outlined in Section~\ref{sec:population} for each timescale. The results for each of these tests is shown in Table~\ref{tab:binom_results}. The most significant result comes when searching for steady neutrino sources over the entire GFU livetime, which yields an analysis $p$-value of 0.018 when comparing the observed $\alpha$ ($6.1\times 10^{-4}$) to a distribution generated from ensembles of pseudo-experiments. We find this to be consistent with expectations from background, especially after considering that we perform three tests, one for each time window, which is not reflected in the quoted $p$-value above (treating the time-windows as independent would result in a trials corrected $p$-value of approximately 5\%). The comparison of the observed value and the background expectation is shown in Figure~\ref{fig:binomial_distribution}, where we also show how we calculate $\alpha$, namely, by calculating $\alpha_k$ for all possible number of sources, $k$, and returning the minimum. The analysis finds a best-fit number of sources of $k=23$ ($\alpha_{23} = 6.1\times 10^{-4}$) for the time-integrated analysis.

\section{Constraints on populations of sources} \label{sec:constraints}

In Section~\ref{sec:intro}, we have described how this analysis is model-independent in that it does not rely on searching for neutrino emission from a specific astrophysical class of objects. Because of this, we can use the results of this analysis to constrain a wide variety of populations of potential astrophysical neutrino sources. 

In order to constrain these populations of possible neutrino sources, we must first simulate how they would appear in the analysis. To do this, we begin by using the publicly available FIRESONG code~\citep{Tung2021}. FIRESONG takes a few variables describing the population as inputs: a neutrino luminosity function characterizing the distribution of intrinsic luminosities of sources, a local number or rate density, a cosmic evolution model as a function of redshift, and an assumed spectral shape for the neutrino emission. Given these inputs, FIRESONG simulates a population of neutrino sources and calculates the neutrino flux at Earth from each of these objects. FIRESONG can simulate either transient or steady neutrino sources. We simulate transient sources when considering the populations we can constrain with the short timescale analyses, and we simulate steady sources when considering populations we can constrain with the time-integrated analysis. Unless stated otherwise, we simulate all populations assuming a spectral index of $\gamma=2.5$, which is consistent with recent measurements of the diffuse astrophysical neutrino flux~\citep{IceCube:2020acn}.

Once we have a list of simulated sources, each with a flux and declination, $\phi_{l}, \delta_l$, we then determine which sources would yield an alert event. In order to do this, we calculate $\langle \mathcal{N}^{\mathrm{alert}}_l(\delta_l, \gamma=2.5, \phi_l)\rangle$ according to Equation~\ref{eq:nevents}. As this number is an expectation, we then randomly generate the actual number of observed alerts by sampling from a Poisson distribution with mean $\langle \mathcal{N}^{\mathrm{alert}}_l(\delta_l, \gamma=2.5, \phi_l)\rangle$ for each simulated source. We are most interested in the regime where $\langle \mathcal{N}^{\mathrm{alert}}_l(\delta_l, \gamma=2.5, \phi_l)\rangle$ is much less than one, because previous limits on neutrino source populations suggest there are more astrophysical neutrino sources than there are observed alert events from astrophysical sources~\citep{IceCube:2018ndw}, and therefore we do not inject more than one alert event when an individual source could yield more than one alert event. However, in these cases, there are many additional events that are still injected into the GFU selection for the first alert, and thus they are still distinguishable from background-like populations in the analysis.

\begin{figure*}
    \centering
    \includegraphics[width=0.6\textwidth]{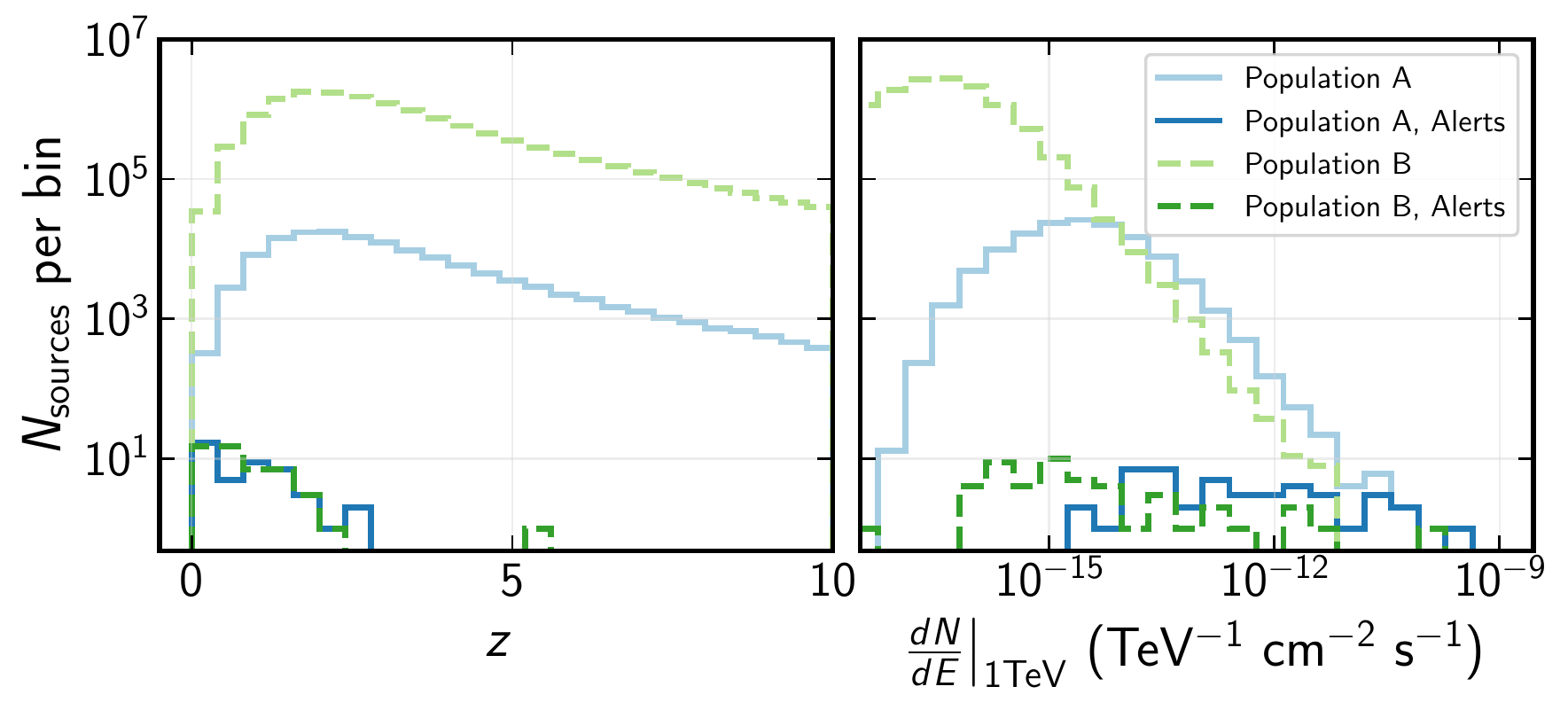}
    \caption{Redshift distribution (left) and flux distribution (right) for two different simulated populations of neutrino sources, both of which saturate the total diffuse astrophysical neutrino flux. The population in green (Population A) has a higher local density ($10^{-6}~\textrm{Mpc}^{-3}$) but a smaller per-source luminosity while the blue population (Population B) has a lower density ($10^{-8}~\textrm{Mpc}^{-3}$) but higher per-source luminosity. Both populations track the star-formation model from \cite{Madau:2014bja} and assume sources have $E^{-2.5}$ spectra. For each population, those sources which result in alert events are shown in the corresponding darker colors. Both populations have the same redshift distribution in alert events, as the cumulative flux distribution as a function of redshift is the same for both populations. Alerts from the rarer population (dark blue) correspond to sources with higher fluxes than the higher density population (dark green).}
    \label{fig:firesong_fluxes}
\end{figure*}

In Figure~\ref{fig:firesong_fluxes}, we show one particular realization 
%of this 
when simulating source populations whose number densities are assumed to track star-formation rates~\citep{Madau:2014bja}. We simulate two different populations, where each of these populations would completely saturate the diffuse neutrino flux. The difference between the two simulated populations is in their number densities and per-source luminosities (explained below). For one population, we simulate a relatively rare population of sources (local number density of $10^{-8}$~Mpc$^{-3}$), and for the other, a relatively high density population of sources (local number density of $10^{-6}$~Mpc$^{-3}$). 

\begin{figure}
    \centering
    \includegraphics[width=0.46\textwidth]{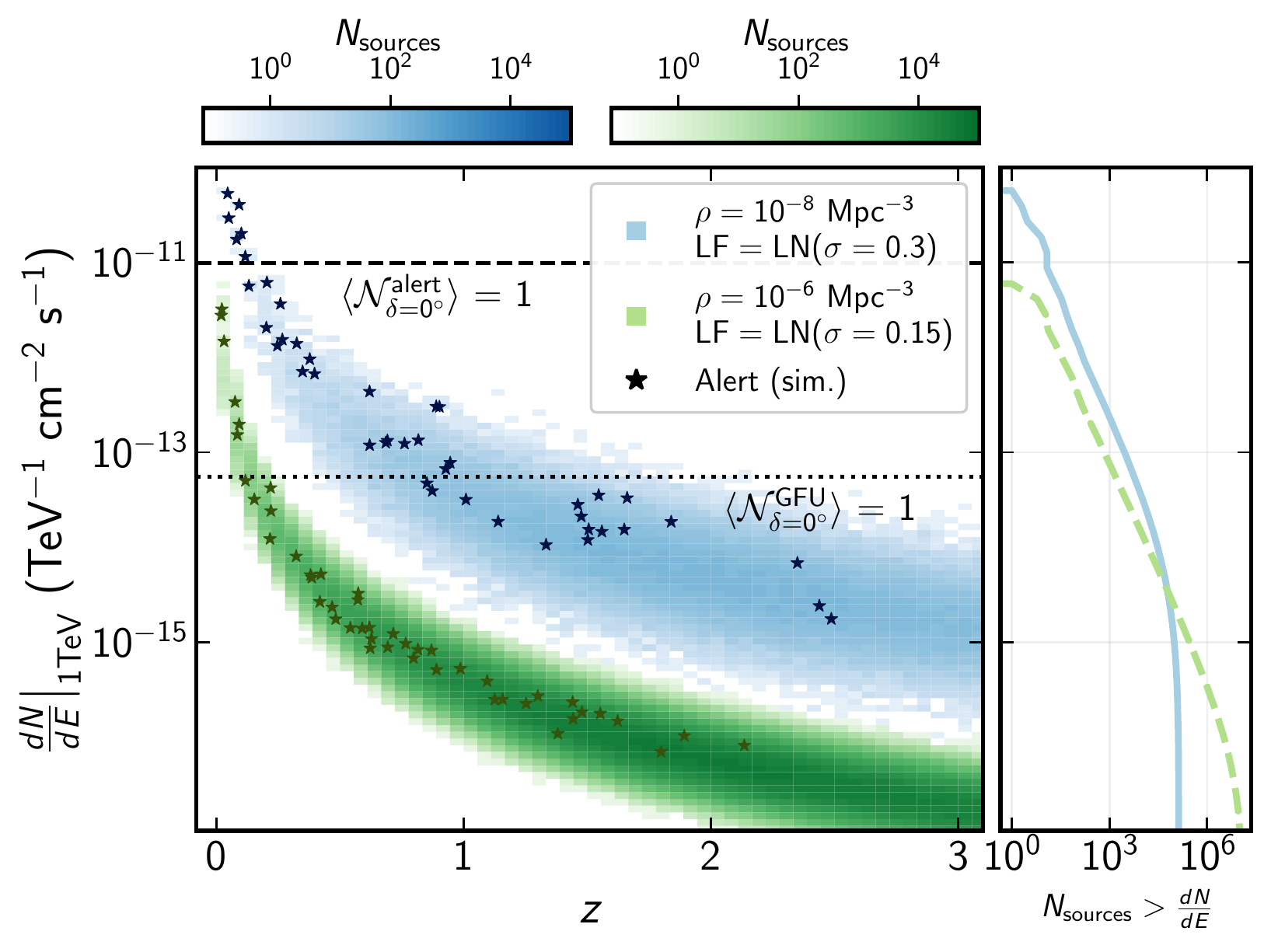}
    \caption{Two dimensional redshift and flux distributions for the same populations of sources shown in Figure~\ref{fig:firesong_fluxes}, whose luminosity functions (LF) follow LogNormal distributions (LN) with widths specified in the legend. The histogram shows the overall flux and redshift distribution, and the stars correspond to those sources that yielded an alert event in a given pseudo-experiment. The population with a smaller density (blue) has alert events that have, on average, higher fluxes. The dashed and dotted black lines show the $E^{-2.5}$ fluxes needed from a source at $\delta=0^{\circ}$ to have an expectation of detecting one alert event or one event in the large-statistics GFU sample, respectively. The higher fluxes from the blue population lead to sources in the upper-left corner of the figure, which produce events in both the alert and the GFU sample. The cumulative $\log(N)-\log(S)$ distribution is given for the low density population (solid blue) and high density population (dashed green) on the right.}
    \label{fig:analysis_motivation_firesong}
\end{figure}

In order for both of these populations to saturate the diffuse flux, which is equivalent to ensuring that the expectation for the best fit flux normalization 
%number of alert events from each population 
matches the observed rate of alerts with astrophysical origin, the population with the smaller number density must have, on average, brighter neutrino sources. This is highlighted in flux distribution shown in Figure~\ref{fig:firesong_fluxes}, as the population with fewer sources extends to higher fluxes. In addition to showing the flux distribution of every source in the population, we also show the subset of sources which resulted in an alert event, for a given realization. 

After we have calculated those sources which will yield alert events in a particular realization, we then find how many events in the GFU sample we expect to observe from those sources, i.e. we calculate $\langle \mathcal{N}^{\mathrm{GFU}}_l(\delta_l, \gamma=2.5, \phi_l)\rangle$ according to Equation~\ref{eq:nevents}, and then fluctuate each of these numbers in the same way that we do for the alert event observations. Once we have a list of sources which caused alert events and we also have the number of additional events in the GFU sample that we should observe in a given pseudo-experiment, we inject these additional events on top of our scrambled data to perform a single pseudo-experiment. In addition to alerts from sources, there are also alerts that are the result of atmospheric backgrounds. For these alert events, we Poisson fluctuate the rates cited in~\cite{Blaufuss:2019fgv} to find the number of expected alerts from atmospheric backgrounds, and we include these in our list of alert events for a psuedo-experiment, but for these alert events, we do not inject any additional events into the GFU sample, and we perform followups for these alerts as well. We calculate all analysis parameters (test statistics as well as stacked binomial $p$-values) to find if the simulated population of sources is distinguishable from our observed data.

The motivation for this analysis lies in the differences between the flux distributions for the sources which cause alert events. Those sources which cause alert events and come from low density source populations have, on average, higher fluxes than sources which cause alert events and come from high density source populations. This means that, although we may be in the Eddington bias regime for alert events (i.e. $\langle \mathcal{N}^{\mathrm{alert}}_l(\delta_l, \gamma, \phi_l)\rangle < 1$), we may not be in the Eddington bias regime for events from the selection with a larger effective area (i.e. $\langle \mathcal{N}^{\mathrm{GFU}}_l(\delta_l, \gamma, \phi_l)\rangle > 1$). If, when we search for sources in the direction of alert events, there are additional GFU events coming from a source, then this allows us to more accurately calibrate the flux of that source. This comparison is made more explicit in Figure~\ref{fig:analysis_motivation_firesong}, where we show the same simulated populations of sources as we did in Figure~\ref{fig:firesong_fluxes}. However, here we also draw attention to the flux normalizations where there is an expectation of observing exactly one event in the alert sample (GFU sample) for a source at $\delta=0^{\circ}$ and with an $E^{-2.5}$ spectrum, i.e. $\langle \mathcal{N}^{\mathrm{alert}}(\delta = 0^{\circ}, \gamma=2.5, \phi)\rangle = 1$ ($\langle \mathcal{N}^{\mathrm{GFU}}(\delta = 0^{\circ}, \gamma=2.5, \phi)\rangle = 1$).

For the populations which we simulate, there is a qualitative difference between the sources which cause alerts in the low-density and high-density populations. Namely, the majority of alert events from the rare population will be accompanied by lower-energy events in the GFU sample, whereas only a small fraction of alert events will have additional detected events when the population is more numerous. From Figure~\ref{fig:analysis_motivation_firesong}, it becomes clear that the some low density populations can be distinguished from high density populations using the population analysis described in Section~\ref{sec:population}. This is because the additional lower-energy GFU events which accompany the alerts in the low density population scenario could be identified in each of the alert followups. This, when stacked together using the population analysis, would be more signal-like than the case of the high density population. 

\begin{figure*}
    \centering
     \includegraphics[width=0.46\textwidth]{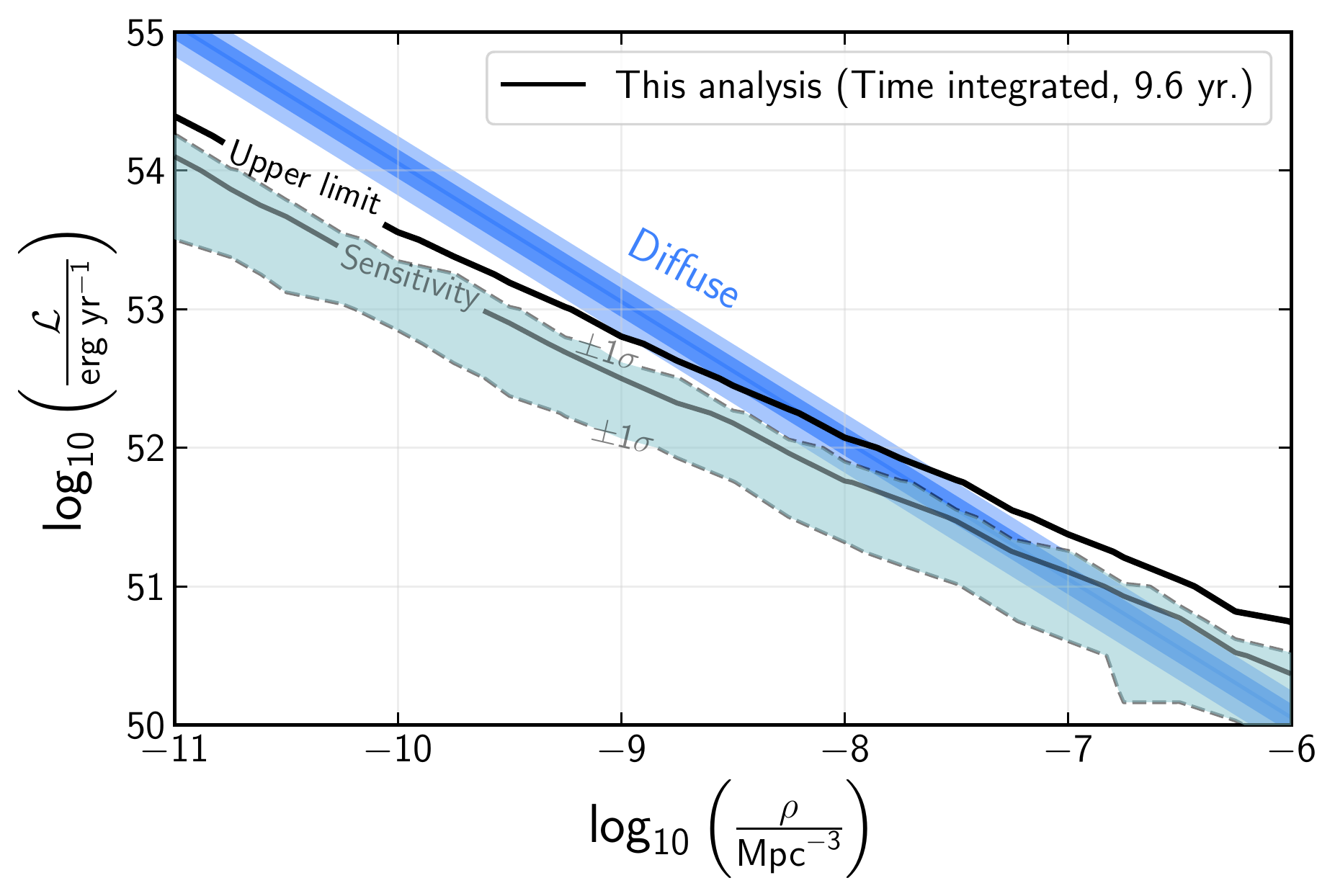}
    \includegraphics[width=0.46\textwidth]{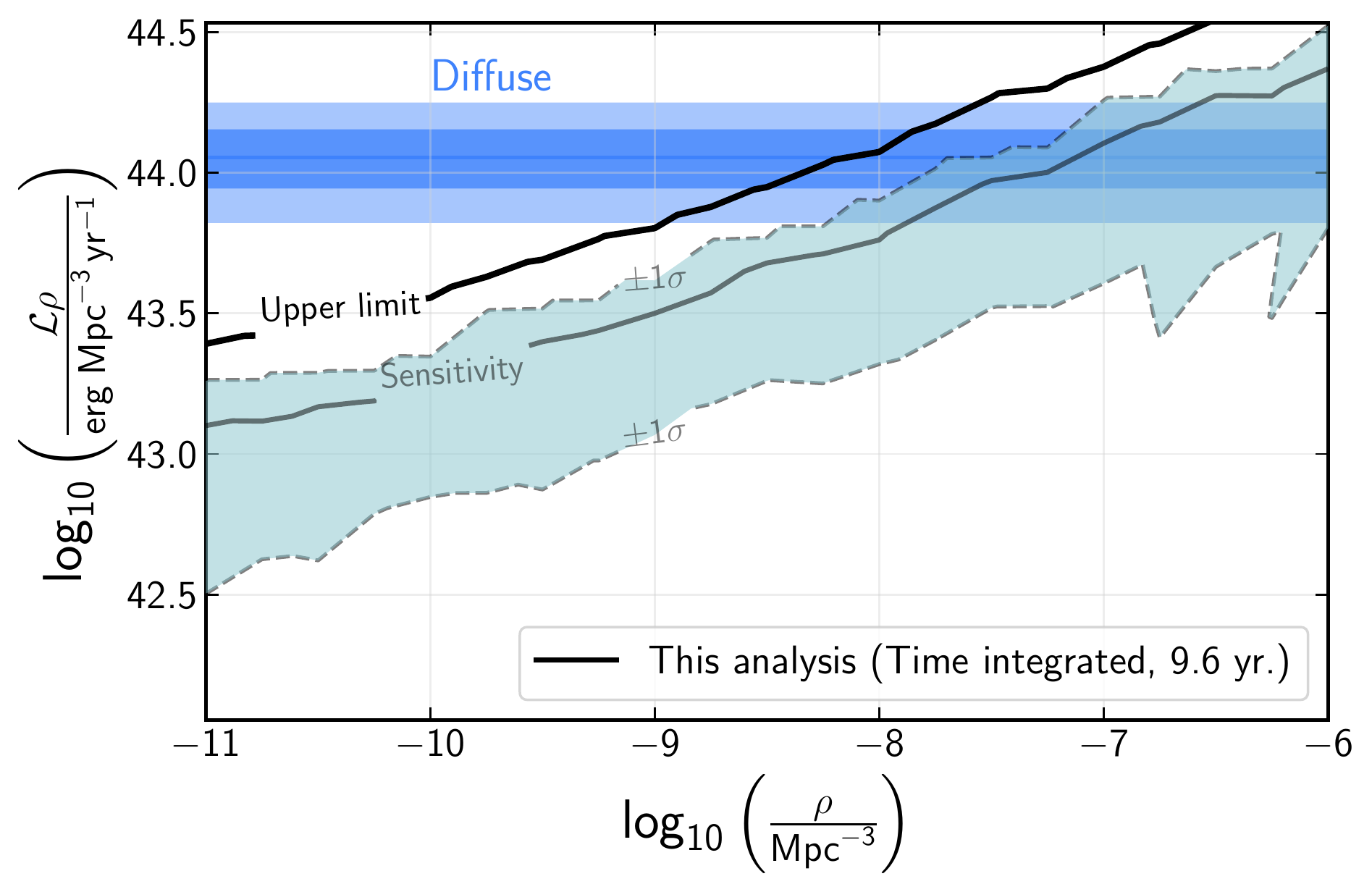}
    \caption{Constraints on the per-source per-flavor neutrino luminosity between 10~TeV and 10~PeV from populations of steady neutrino sources with $E^{-2.5}$ spectra, and whose densities track star-formation rates. The faded line and band show the analysis sensitivity and the 1$\sigma$ (68\%) expected fluctuation of a one-sided Neyman upper-limit under the null hypothesis. The data are inconsistent with a sole population of sources with the same luminosity being responsible for the diffuse flux (shown with uncertainties as the blue shaded regions) unless it has a density greater than $7\times 10^{-9}~\textrm{Mpc}^{-3}$. The left hand side shows these constraints in the density/luminosity plane, as in~\cite{Kowalski:2014zda}, whereas the figure on the right scales the luminosity by density, which is proportional to the energy density, to focus on the most-relevant section of the parameter space.}
    \label{fig:steady_upper_limits}
\end{figure*}

In order to quantify how distinguishable certain populations are from our observed data, we repeat this process of injecting source populations into the analysis for a variety of different source input parameters. For each of the spots in the input source parameter space, we calculate expected distributions of binomial $p$-values, $\alpha$, and we compare this to our observed values for both steady source hypotheses as well as transient source hypotheses. 

\begin{figure*}
    \centering
    \includegraphics[width=0.46\textwidth]{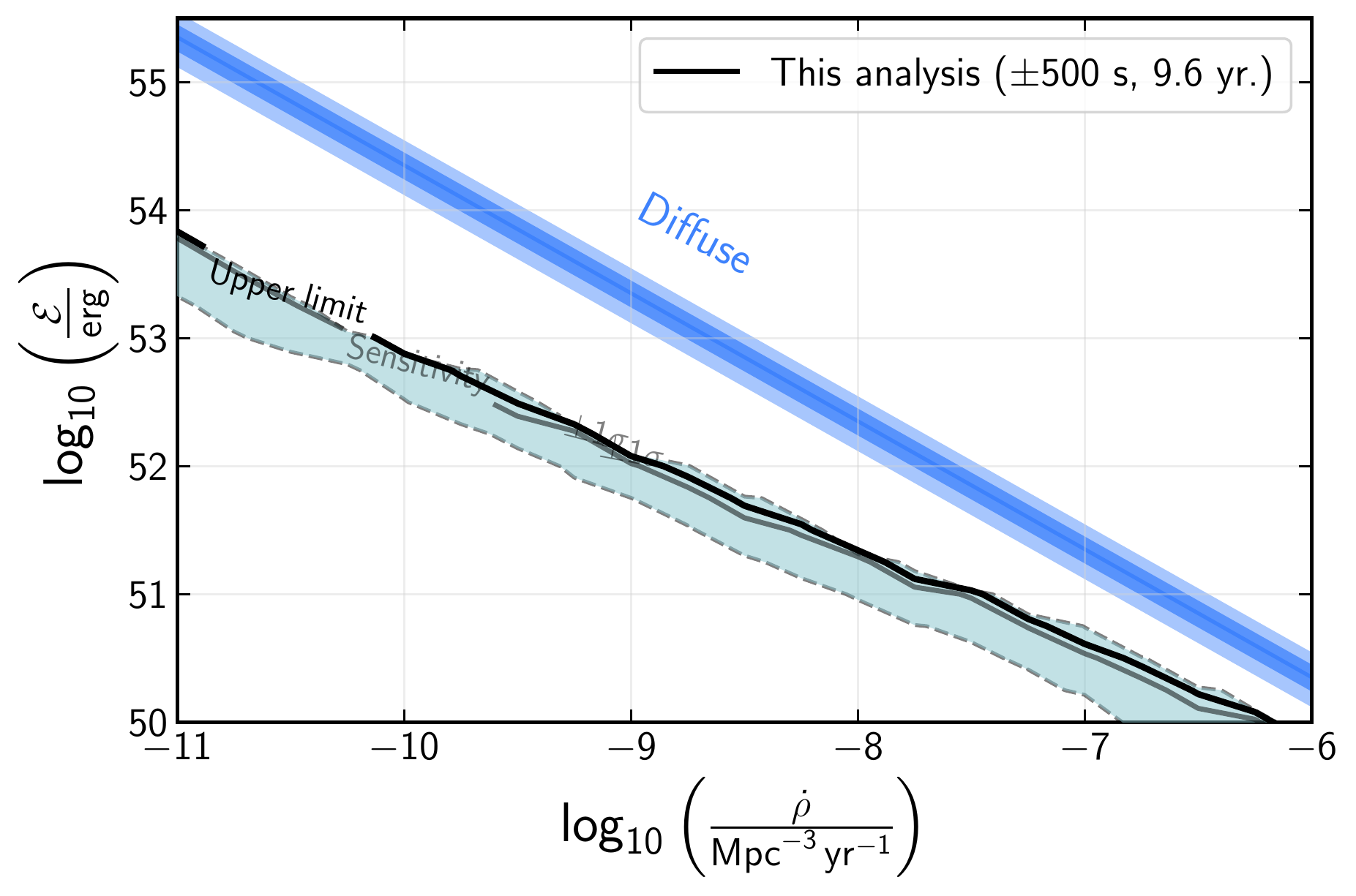}
    \includegraphics[width=0.46\textwidth]{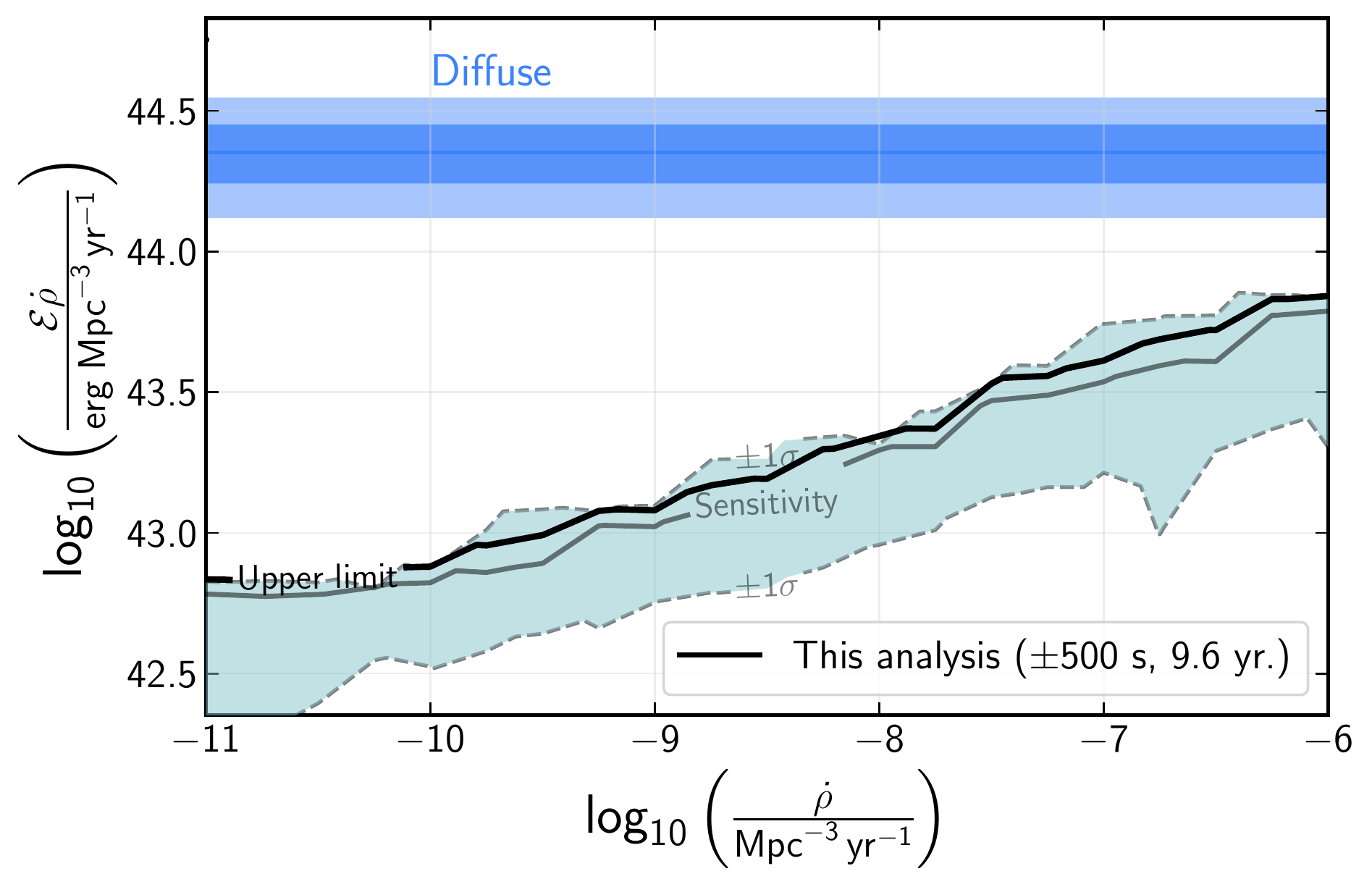}
    \caption{Constraints on the per-source emitted energy per-flavor between 10~TeV and 10~PeV from populations of transient neutrino sources with $E^{-2.5}$ spectra whose rate densities track star-formation rates. The faded line and band show the analysis sensitivity and the 1$\sigma$ (68\%) expected fluctuation of a one-sided Neyman upper-limit under the null hypothesis. Upper limits (90\% CL) are inconsistent with rare populations (rate density less than $10^{-9}$~Mpc$^{-3}$~yr$^{-1}$), of standard candle transients producing more than 6\% of the diffuse flux (shown with uncertainties as the blue shaded regions) for minute-scale  transients.}
    \label{fig:transient_upper_limits_1}
\end{figure*}

\begin{figure*}
    \centering
    \includegraphics[width=0.46\textwidth]{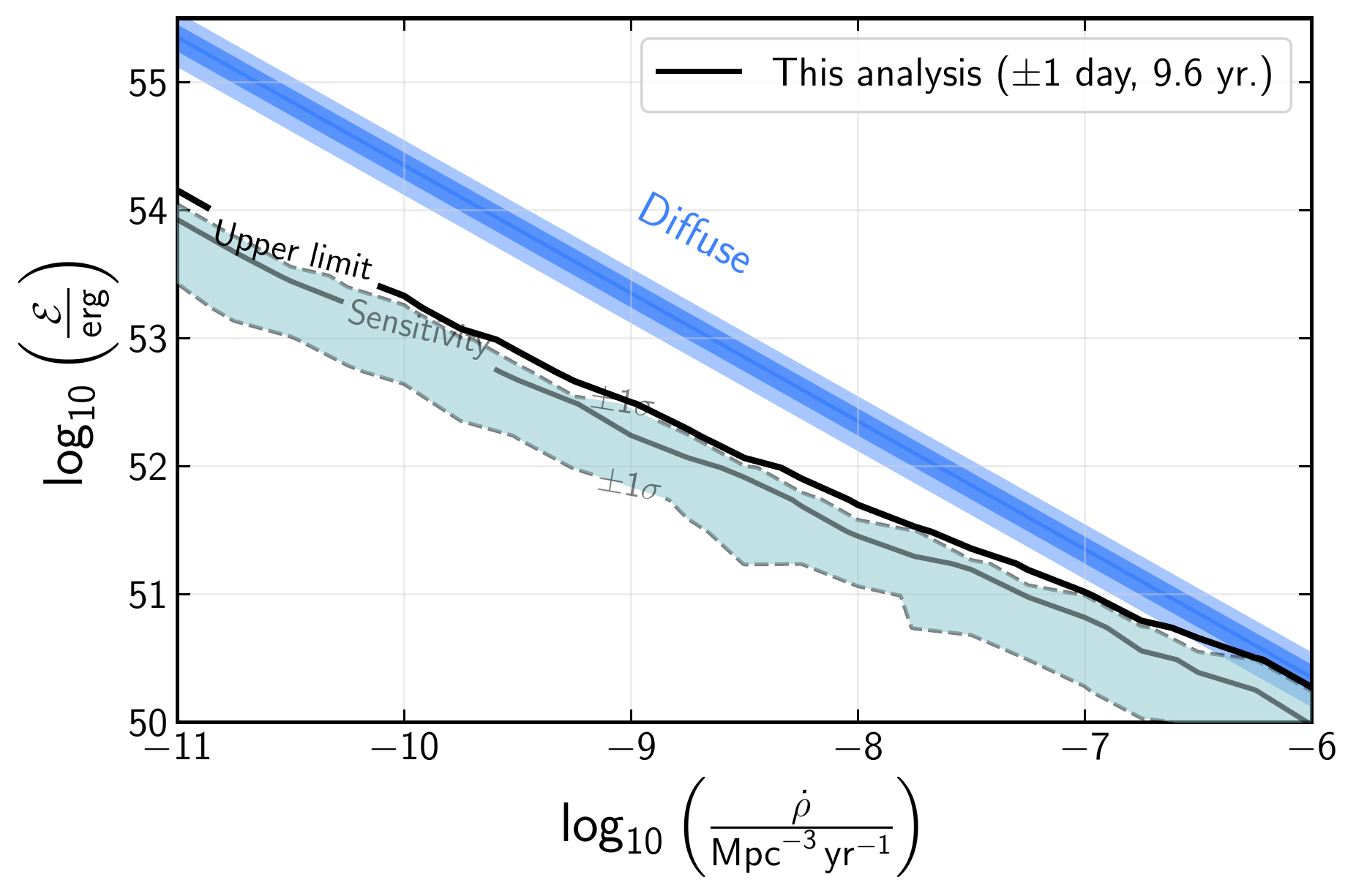}
    \includegraphics[width=0.46\textwidth]{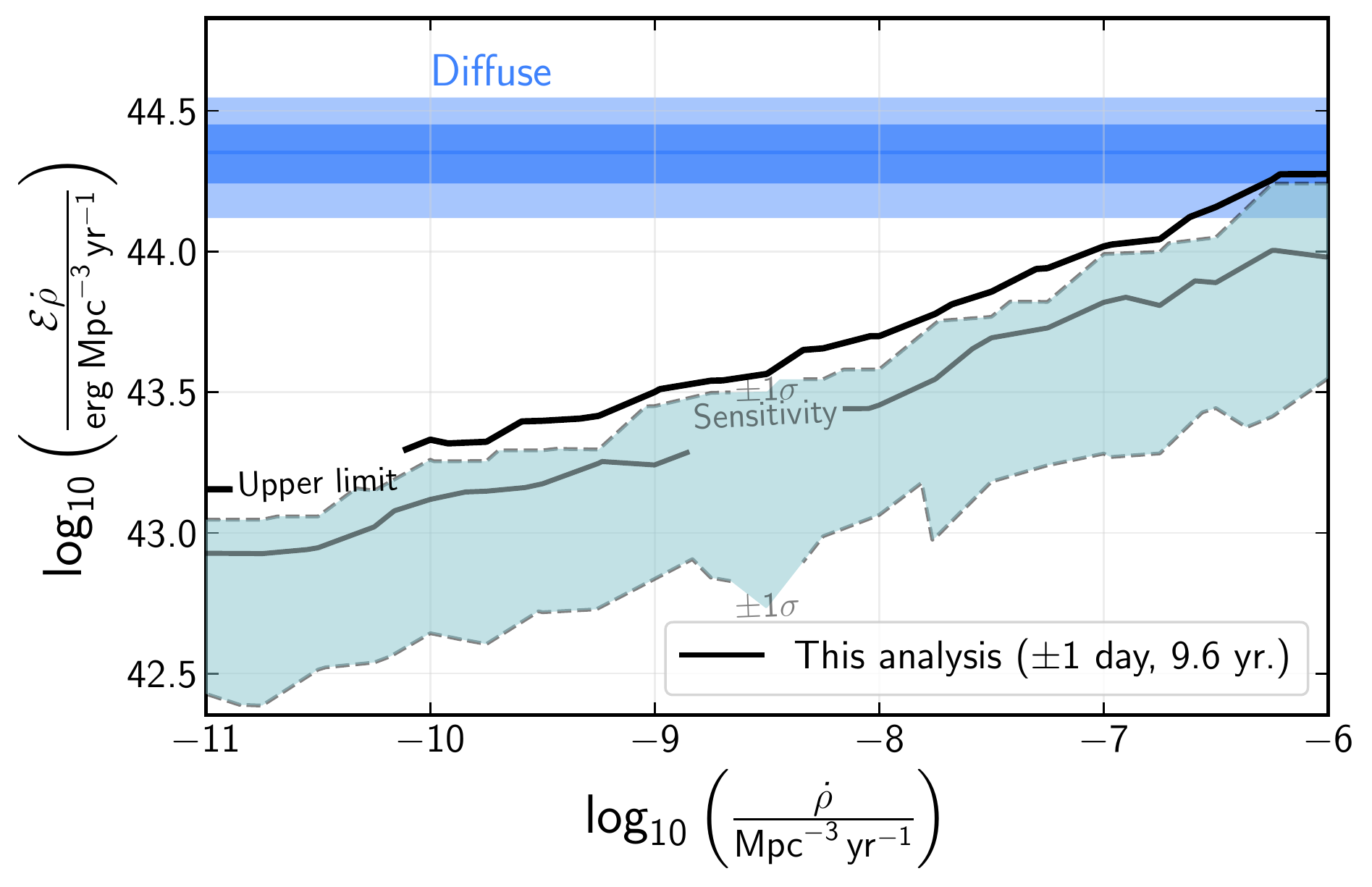}
    \caption{Constraints on the per-source emitted energy per-flavor between 10~TeV and 10~PeV from populations of transient neutrino sources with $E^{-2.5}$ spectra whose rate densities track star-formation rates. The faded line and band show the analysis sensitivity and the 1$\sigma$ (68\%) expected fluctuation of a one-sided Neyman upper-limit under the null hypothesis. Upper limits (90\% CL) are inconsistent with rare populations (rate density less than $10^{-9}$~Mpc$^{-3}$~yr$^{-1}$), of standard candle transients producing more than 14\% of the diffuse flux (shown with uncertainties as the blue shaded regions) for day-scale transients.}
    \label{fig:transient_upper_limits_2}
\end{figure*}

Our resulting per-flavor limits on steady neutrino source populations are shown in Figure~\ref{fig:steady_upper_limits}. Our per-flavor limits on transient source populations, which we calculate based off of our results from the short-timescale analyses, are displayed in Figures~\ref{fig:transient_upper_limits_1} and ~\ref{fig:transient_upper_limits_2}. For the transient analysis, we inject populations of transients where the emission timescale matches the analysis duration in the observer frame, and thus the $\pm500$~s results can be used to constrain any transient sources where the true emission in the observer frame is less than or equal to $500$~s, and the $\pm 1$~day timescale can be used to constrain any transient source population with true durations in the observer frame less than or equal to 1~day (we will call these minute-scale and day-scale transients, respectively). These limits are compared to the median upper limits that one could expect 
%to set 
under the assumption of the null-hypothesis. In Figures~\ref{fig:steady_upper_limits}, ~\ref{fig:transient_upper_limits_1} and ~\ref{fig:transient_upper_limits_2}, the faded band represents how much the upper limit (90\% CL) can typically fluctuate under the assumption of the null hypothesis. In case of an underfluctuation, previous analyses have quoted the sensitivity as the upper limit in these cases \citep[see for e.g. ][]{ulirg_stacking}. However, the binomial analyses for all three time windows analyzed here results in pre-trial p-values that are less than p=0.5, and thus we quote upper limits that are calculated using the classical Neyman approach.

The input parameter space of potential source populations is highly-dimensional, so we choose some benchmark models for these limits. Namely, we assume that source densities track star-formation rates as measured in~\cite{Madau:2014bja} and we assume that all sources have a ``standard candle'' (SC) luminosity function, i.e. one in which all of the sources have the same luminosity and the observed flux from an individual source is solely determined by its distance. Under these assumptions, the time-integrated analysis shows that if a population of standard candle sources whose densities track star-formation rates were to be solely responsible for the diffuse astrophysical neutrino flux, then their local density must be greater than $7\times 10^{-9}$~Mpc$^{-3}$ at the 90\% confidence level. For this time-integrated analysis, the fact that our observed result was more signal-like than our median expectation from the null hypothesis is reflected in the weakening of our limits with respect to the median expectation, as shown in Figure~\ref{fig:steady_upper_limits}. For transient source populations, we find that rare populations of transient sources (rate density less than $10^{-9}$~Mpc$^{-3}$~yr$^{-1}$), can be responsible for no more than 6\% (14\%) of the diffuse flux for minute-scale (day-scale) transients. We also extrapolate these limits to higher densities in the same manner as~\cite{IceCube:2018ndw}. This tells us that populations of transient sources with these same population parameters cannot be solely responsible for the diffuse flux unless their local rate densities are greater than $8\times 10^{-5}$~Mpc$^{-3}$~yr$^{-1}$ ($1\times 10^{-5}$~Mpc$^{-3}$~yr$^{-1}$) for minute-scale (day-scale) transient sources.
%Extrapolating these limits to higher densities in the same manner as~\cite{IceCube:2018ndw} tells us that populations of transient sources with these same population parameters cannot be solely responsible for the diffuse flux unless their local rate densities are greater than $8\times 10^{-5}$~Mpc$^{-3}$~yr$^{-1}$ ($1\times 10^{-5}$~Mpc$^{-3}$~yr$^{-1}$) for minute-scale (day-scale) transient sources.

\begin{table*}
    \centering
    \begin{tabular}{|c | c | c c | c c | c c|} \cline{3-8} \cline{3-8}
         \multicolumn{2}{c|}{} & \multicolumn{6}{c|}{Fraction of diffuse flux} \\ \cline{3-8}
        \multicolumn{2}{c|}{} &  \multicolumn{2}{c|}{$10^{-10}$~Mpc$^{-3}$ (Mpc$^{-3}$~yr$^{-1}$)} & \multicolumn{2}{c|}{$10^{-8}$~Mpc$^{-3}$ (Mpc$^{-3}$~yr$^{-1}$)} & \multicolumn{2}{c|}{$10^{-6}$~Mpc$^{-3}$ (Mpc$^{-3}$~yr$^{-1}$)} \\ \cline{3-8}
        \multicolumn{2}{c|}{} &  \textbf{SC} & \textbf{LN}, $\sigma = 0.4$ & \textbf{SC} & \textbf{LN}, $\sigma = 0.4$ & \textbf{SC} & \textbf{LN}, $\sigma = 0.4$ \\ \hline
        \multirow{2}{*}{$\mathbf{\pm 500}$~\textbf{s}} & SFR & 0.036 & 0.033 & 0.10 & 0.090 & 0.32 & 0.25 \\
        & No Evolution & 0.028 & 0.032 & 0.037 & 0.035 & 0.11 & 0.11 \\ \hline
        \multirow{2}{*}{$\mathbf{\pm 1}$~\textbf{day}} & SFR & 0.092 & 0.083 & 0.23 & 0.21 & 0.91 & 0.66 \\
        & No Evolution & 0.062 & 0.065 & 0.091 & 0.082 & 0.26 & 0.27 \\  \hline
        \multirow{2}{*}{\textbf{Time-integrated}} & SFR & 0.32 & 0.30 & 1.10 & 0.94 & 4.60 & 3.67 \\
        & No Evolution & 0.28 & 0.33 & 0.44 & 0.40 & 1.41 & 1.18 \\ \hline
    \end{tabular}
    \caption{Constraints on the fractional contribution to the diffuse astrophysical neutrino flux from populations of steady (transient) neutrino sources  for various densities (rate densities). Populations either track star-formation rates (SFR) or have no cosmic evolution (No Evolution). For each population, we include how the limits change for standard-candle (SC) populations, in which all sources have the same luminosity, as well as for LogNormal (LN) luminosity functions with characteristic width of $\sigma=0.4$. All populations assume a diffuse astrophysical flux with spectral index $\gamma=2.5$.
    } \label{tab:populations}
\end{table*}

We explore how our limits change by altering these assumptions, and the results of that analysis are tabulated in Table~\ref{tab:populations}. Overall, we find that our limits are not extremely sensitive to changes in the assumed luminosity function. However, changes in the cosmic evolution can have a large effect on the limits. In general, the more rapidly the number of sources grows as a function of redshift, the less constraining this analysis will be on the local density, as rapidly growing source populations have a larger total number of sources at higher redshifts that could result in alert events. 

Although we choose models as generic as possible for source evolution (to reflect the fact that we do not know the true properties of the source classes responsible for the diffuse flux), we also inject a more rapidly evolving population into the time-integrated analysis, to see how these limits might change for even more strongly evolving populations. For this example, we inject a toy-model that was used to describe a generic AGN-like evolution, in which the density of sources $\rho(z)$ is described as a piecewise function of $z$, 
\begin{equation}
   \rho(z) = \rho_0 \times \begin{cases}
           (1 + z)^5 \quad & z \leq z_a \\
           (1 + z_a)^5 & z_a < z \leq z_b \\
           (1 + z_a)^5 \cdot 10^{z_b - z} & z > z_b
        \end{cases}
\end{equation}
where $z_a=1.7$ and $z_b=2.7$. The observational study underlying this parameterization is originally from~\cite{Hasinger:2005sb} and was later reduced to this one-dimensional simplification in~\cite{Stanev:2008un}. For this case, treating the luminosity function again as a standard candle population, we find we can only exclude populations rarer than $6\times 10^{-11}$~Mpc$^{-3}$, which is nearly two orders of magnitude less numerous than the local density of sources we can exclude if sources are assumed to follow star-formation rates. This is due to the greater number of sources at higher redshifts $z \gtrsim 1$ for the AGN-like evolution compared to the star-formation-like evolution. Although the luminosity function of AGN is known to not be a standard candle function, we treated the luminosity function as such for this example because we know that the effect from varying luminosity functions is second-order when compared to the density evolution.  

These limits are the first limits reported by IceCube on populations of sources from searches for neutrino sources in the direction of alert events. Other population constraints, such as the ones reported in~\cite{IceCube:2018ndw}, are at a similar level for standard candle sources, although that analysis constrained sources with harder spectra. Although other neutrino source analyses tend to suffer when searching for neutrino sources with soft spectra, this analysis does not suffer as much because the ratio $\langle \mathcal{N}^{\mathrm{GFU}}_l(\delta, \gamma, \phi)\rangle / \langle \mathcal{N}^{\mathrm{alert}}_l(\delta, \gamma, \phi)\rangle$ gets larger for soft spectra, as the alert stream is most sensitive at higher energies, which effectively boosts our signal when looking for correlations with alerts by using lower-energy events. When we inject harder spectra into our analysis, the constraints get weaker than those reported in~\cite{IceCube:2018ndw}. As for limits on transient source populations, the limits on the total emitted energy agree within 10\% with the limits reported in~\cite{IceCube:2018omy}. However, those limits were only on sources with emission timescales less than 100~s in duration, meaning that it was effectively insensitive to transients that emit on longer timescales. Thus, the limits on transient source populations presented here are the first limits reported by IceCube on transient source populations with emission timescales greater than 100~s and up to one day in duration.

\section{Discussion \& Conclusion}
\label{sec:conclusion}

We present a search for neutrino events in the direction of IceCube neutrino candidate alert events. This search strategy is model-independent in that it does not rely on any assumptions of a specific source class being responsible for the diffuse neutrino flux. 
However, it is complementary to other model-independent searches that are typically  accompanied by large trials-factor, because we only need to search for emission coincident with the smaller statistics sample of alert quality events. 
%However, it is able to avoid the pitfall of a large trials-factor that typically accompanies other model-independent searches, because we only need to search for emission coincident with the smaller statistics sample of alert quality events. 

We search for neutrino emission on short timescales coincident with the alert events as well as for emission over the entire livetime of our datasets. No individual followup yields a significant result. We use these results to constrain contributions to the diffuse astrophysical neutrino flux from generic populations of sources.

Our most significant result comes from the time-integrated population analysis, where we look for joint contributions from multiple subthreshold sources in the direction of alert events. Although consistent with background ($p$-value of 1.8\% before accounting for the three time windows investigated), it cannot be said that there are no sources in the direction of these alert events. This is merely an indication that any sources that are in the directions of these alert events are currently at a level that is too dim to be significantly detected with the current analysis sensitivity. As of July 2021, we began using the procedure described in Section~\ref{sec:analysis} to search for coincident transient neutrino emission for alerts that were detected in real time, with results frequently circulated via the Gamma-ray Coordinates Network in the hopes of identifying future sources that are producing alerts.

Future improvements might begin to reveal sources that were subthreshold to this analysis. One major improvement could come from a better localization of IceCube alert events. Reducing the localization uncertainty on these events would reduce the effective background of this analysis and limit the number of locations on the sky that we need to investigate. Studies are underway to find the optimal way to construct these localization uncertainty spaces~\cite{IceCube:2021uwf}. Additionally, if sources are active in the time-domain and may emit lower-energy neutrinos at different times than we detect high-energy alert events. This would mean, an analysis that can search for this, similar as to what was done for TXS~0506+056, would be more sensitive to these fluxes. Such an analysis is in progress, with preliminary sensitivities reported in~\cite{IceCube:2021kxg}. Finally, a better treatment of the point-source likelihood could improve the per-source sensitivity to neutrino sources. Improvements to this analysis are underway, and a population analysis that utilizes these changes could prove fruitful in identifying populations of individually subthreshold neutrino sources~\citep{Bellenghi:2021uef}.

\section*{Acknowledgements}
The IceCube collaboration acknowledges the significant contributions to this manuscript from Alex Pizzuto, Abhishek Desai and Justin Vandenbroucke. The authors gratefully acknowledge the support from the following agencies and institutions: 
USA {\textendash} U.S. National Science Foundation-Office of Polar Programs,
U.S. National Science Foundation-Physics Division,
U.S. National Science Foundation-EPSCoR,
Wisconsin Alumni Research Foundation,
Center for High Throughput Computing (CHTC) at the University of Wisconsin{\textendash}Madison,
Open Science Grid (OSG),
Advanced Cyberinfrastructure Coordination Ecosystem: Services {\&} Support (ACCESS),
Frontera computing project at the Texas Advanced Computing Center,
U.S. Department of Energy-National Energy Research Scientific Computing Center,
Particle astrophysics research computing center at the University of Maryland,
Institute for Cyber-Enabled Research at Michigan State University,
and Astroparticle physics computational facility at Marquette University;
Belgium {\textendash} Funds for Scientific Research (FRS-FNRS and FWO),
FWO Odysseus and Big Science programmes,
and Belgian Federal Science Policy Office (Belspo);
Germany {\textendash} Bundesministerium f{\"u}r Bildung und Forschung (BMBF),
Deutsche Forschungsgemeinschaft (DFG),
Helmholtz Alliance for Astroparticle Physics (HAP),
Initiative and Networking Fund of the Helmholtz Association,
Deutsches Elektronen Synchrotron (DESY),
and High Performance Computing cluster of the RWTH Aachen;
Sweden {\textendash} Swedish Research Council,
Swedish Polar Research Secretariat,
Swedish National Infrastructure for Computing (SNIC),
and Knut and Alice Wallenberg Foundation;
European Union {\textendash} EGI Advanced Computing for research;
Australia {\textendash} Australian Research Council;
Canada {\textendash} Natural Sciences and Engineering Research Council of Canada,
Calcul Qu{\'e}bec, Compute Ontario, Canada Foundation for Innovation, WestGrid, and Compute Canada;
Denmark {\textendash} Villum Fonden, Carlsberg Foundation, and European Commission;
New Zealand {\textendash} Marsden Fund;
Japan {\textendash} Japan Society for Promotion of Science (JSPS)
and Institute for Global Prominent Research (IGPR) of Chiba University;
Korea {\textendash} National Research Foundation of Korea (NRF);
Switzerland {\textendash} Swiss National Science Foundation (SNSF);
United Kingdom {\textendash} Department of Physics, University of Oxford.

\appendix
% \onecolumngrid
\section{Individual Followup Results Table}
\label{app:results_table}
\onecolumngrid
\newpage
\begin{longdeluxetable}{lllccllc}
\tabletypesize{\scriptsize}
\tablecaption{Results from the individual skymap analyses for each of the time windows analyzed. In addition to providing the best-fit position from the alert event skymaps in Equatorial coordinates (J2000, all coordinates quoted in degrees), we also provide the location which yielded the largest test-statistic for the time-integrated analysis in the (RA, dec.)$_{\mathrm{steady}}$ column. For the short time windows, we only provide the $p$-values from each analysis, and for the time-integrated analysis, we also provide the test-statistic and the best-fit parameters from the likelihood analysis. Those alerts which were excluded from the analyses, for reasons described in the text, are labeled with ``Excl.'' in the relevant columns. Times are quoted in terms of Modified Julian Day (MJD). \label{tab:results}}

\tablehead{\colhead{RunID:EventID} &     \colhead{(RA, dec.)} &      \colhead{MJD} & \colhead{$-\log_{10}(p_{\pm500\mathrm{s}})$} & \colhead{$-\log_{10}(p_{\pm1\mathrm{day}})$} & \colhead{(RA, dec.)$_{\mathrm{steady}}$} & \colhead{(TS, $n_s$, $\gamma$)$_{\mathrm{steady}}$} & \colhead{$-\log_{10}(p_{\mathrm{steady}})$}}

\startdata
118178:17334444 &  138.47, -1.94 & 55695.06 &                               0.00 &                               0.82 &                  139.63, -1.94 &                         (2.37, 3.6, 2.14) &                              0.09 \\
118309:46569873 & 272.55, +35.64 & 55722.43 &                               0.00 &                               0.00 &                 272.63, +36.50 &                         (1.06, 5.9, 2.29) &                              0.21 \\
118342:24578488 &   71.15, +5.38 & 55728.73 &                               3.22 &                               1.42 &                   71.40, +4.85 &                        (0.83, 25.4, 3.65) &                              0.18 \\
118435:58198553 &  68.20, +40.67 & 55756.11 &                               0.00 &                               0.00 &                             -- &                            (0.00, --, --) &                              0.00 \\
118475:52691508 &  151.08, +6.99 & 55768.51 &                               0.00 &                               0.00 &                             -- &                            (0.00, --, --) &                              0.00 \\
118539:54350726 &  336.80, +1.53 & 55780.98 &                               0.00 &                               1.55 &                  337.05, +1.92 &                        (1.34, 20.5, 3.04) &                              0.22 \\
118576:45696357 &  332.45, -2.09 & 55791.69 &                               0.00 &                               0.00 &                  332.58, -2.09 &                         (0.23, 8.6, 2.76) &                              0.17 \\
118631:36844560 &    9.76, +7.59 & 55806.09 &                               0.00 &                               0.00 &                    9.37, +7.57 &                         (4.56, 2.2, 1.80) &                              0.81 \\
118660:61529737 &  196.08, +9.40 & 55811.79 &                               0.00 &                               0.00 &                  195.57, +9.54 &                        (3.13, 31.3, 4.08) &                              0.32 \\
  118738:995844 & 121.45, +50.04 & 55833.26 &                               2.48 &                               1.87 &                 120.83, +50.17 &                        (5.93, 40.8, 3.82) &                              1.41 \\
118741:43101116 &  267.01, -4.44 & 55834.45 &                               0.00 &                               0.67 &                  266.55, -5.07 &                        (5.32, 46.7, 3.50) &                              1.20 \\
118778:61465314 & 172.13, +44.70 & 55846.87 &                               0.00 &                               0.00 &                             -- &                            (0.00, --, --) &                              0.00 \\
118973:22324184 &   26.06, +9.82 & 55885.96 &                               0.00 &                               0.00 &                   26.47, +9.29 &                         (4.17, 4.4, 1.85) &                              0.61 \\
118973:25391094 & 356.84, -11.99 & 55885.97 &                               0.00 &                               0.00 &                             -- &                            (0.00, --, --) &                              0.00 \\
119097:49522606 & 165.19, +38.49 & 55903.72 &                               0.00 &                               0.42 &                 165.69, +38.62 &                        (5.05, 31.9, 3.14) &                              0.27 \\
119101:32538906 &  99.98, +20.42 & 55904.46 &                               0.00 &                               0.65 &                  99.84, +20.46 &                         (0.43, 2.4, 2.06) &                              0.11 \\
119125:14422170 &  247.85, +0.56 & 55908.40 &                               0.00 &                               0.00 &                  247.19, +0.60 &                        (0.81, 19.9, 3.56) &                              0.10 \\
119136:66932419 &  36.74, +18.88 & 55911.28 &                               0.00 &                               0.66 &                  37.02, +16.12 &                        (2.15, 70.5, 3.61) &                              0.15 \\
119146:28561103 &   26.85, +7.03 & 55913.34 &                               0.00 &                               0.00 &                   24.54, +7.17 &                        (8.14, 45.7, 2.94) &                              0.62 \\
119739:41603205 & 237.96, +18.76 & 55987.81 &                               0.00 &                               0.00 &                 237.83, +18.84 &                        (1.27, 26.3, 3.90) &                              0.57 \\
120027:12133428 &  183.56, +0.52 & 56043.42 &                               0.00 &                               0.00 &                  184.38, +0.90 &                        (5.99, 40.7, 2.92) &                              0.88 \\
120045:22615214 & 165.37, -71.51 & 56048.57 &                               0.00 &                               0.00 &                 164.64, -70.39 &                       (11.97, 14.1, 2.69) &                              1.45 \\
120157:59255331 & 198.94, +32.00 & 56062.96 &                               0.00 &                               1.10 &                 199.28, +31.45 &                        (0.97, 17.4, 4.90) &                              0.28 \\
120185:63879322 & 171.08, +26.44 & 56070.57 &                               0.00 &                               0.00 &                 170.15, +26.56 &                        (8.33, 57.4, 3.78) &                              1.97 \\
 120186:1306178 & 343.78, +15.48 & 56070.64 &                               0.00 &                               0.56 &                 345.39, +14.82 &                        (1.37, 33.3, 4.54) &                              0.03 \\
120205:73761892 & 176.48, +22.87 & 56076.54 &                               0.00 &                               0.45 &                 177.90, +23.19 &                        (7.77, 35.6, 2.80) &                              0.71 \\
120244:21476686 & 119.31, +14.79 & 56079.31 &                               0.00 &                               1.89 &                             -- &                            (0.00, --, --) &                              0.00 \\
120260:65493286 & 152.58, +36.38 & 56083.65 &                               0.00 &                               0.00 &                 152.94, +36.37 &                         (4.27, 7.5, 2.02) &                              0.70 \\
120309:20451977 &  39.95, -15.09 & 56089.36 &                               0.00 &                               0.00 &                             -- &                            (0.00, --, --) &                              0.00 \\
120535:24248640 &  330.07, +1.42 & 56146.21 &                               0.00 &                               0.00 &                  329.77, +1.49 &                        (0.48, 17.7, 4.72) &                              0.24 \\
120680:12284754 &  182.24, +3.88 & 56186.31 &                               0.00 &                               0.60 &                  181.48, +4.20 &                        (0.81, 29.4, 3.54) &                              0.27 \\
120708:53550535 &  70.62, +19.79 & 56192.55 &                               0.00 &                               0.00 &                  70.28, +19.93 &                        (2.52, 22.8, 3.11) &                              0.60 \\
120798:20915945 &  205.14, -2.28 & 56211.77 &                               0.00 &                               1.28 &                             -- &                            (0.00, --, --) &                              0.00 \\
120860:64513030 & 169.80, +27.91 & 56226.60 &                               0.00 &                               0.00 &                 169.95, +27.37 &                        (7.54, 55.3, 4.33) &                              1.51 \\
120884:64276946 &  123.18, +6.05 & 56234.51 &                               0.00 &                               1.84 &                  124.09, +6.30 &                        (3.63, 55.6, 3.52) &                              0.87 \\
120935:29325450 &  225.70, +8.88 & 56246.33 &                               1.85 &                               0.00 &                  226.06, +8.32 &                        (3.89, 42.6, 3.15) &                              0.96 \\
121761:41853263 &   7.67, +74.14 & 56317.27 &                               0.00 &                               0.00 &                   9.49, +72.86 &                        (9.09, 61.0, 3.69) &                              2.19 \\
121762:56963417 &  280.46, -1.90 & 56317.66 &                               0.00 &                               0.72 &                  280.09, -1.64 &                        (4.10, 23.9, 2.88) &                              0.56 \\
121769:40349510 &  352.97, -1.98 & 56319.28 &                               0.00 &                               0.00 &                  352.62, -1.94 &                        (0.11, 10.3, 2.85) &                              0.12 \\
121840:62872761 &  48.38, -13.32 & 56331.12 &                               0.00 &                               0.00 &                  48.12, -13.30 &                        (1.73, 10.8, 3.29) &                              0.63 \\
 122055:5809789 & 303.41, +54.68 & 56367.74 &                               0.00 &                               0.84 &                 304.23, +53.05 &                        (3.95, 48.7, 4.30) &                              0.51 \\
122060:56194427 &  13.45, +20.62 & 56369.28 &                               0.00 &                               1.60 &                  13.71, +21.57 &                        (1.76, 12.4, 2.43) &                              0.50 \\
122152:30701331 & 167.83, +20.66 & 56390.19 &                               0.00 &                               0.00 &                 165.67, +21.28 &                        (3.39, 38.6, 3.22) &                              0.45 \\
 122154:6905361 &    7.38, +4.22 & 56390.76 &                               0.00 &                               0.48 &                    8.02, +3.88 &                       (13.39, 70.0, 3.61) &                              1.46 \\
122160:52726834 & 163.56, +29.44 & 56391.98 &                               0.00 &                               0.00 &                 163.47, +27.93 &                        (9.31, 29.7, 2.60) &                              1.26 \\
122316:49995489 & 337.76, +26.24 & 56420.64 &                               0.00 &                               1.15 &                 338.17, +26.00 &                        (0.97, 14.0, 3.73) &                              0.07 \\
122318:21601607 &  317.50, +2.09 & 56421.19 &                               0.00 &                               0.00 &                  317.76, +2.18 &                         (0.57, 1.6, 1.99) &                              0.08 \\
122361:46565711 &  45.35, +23.85 & 56431.48 &                               0.00 &                               0.00 &                  46.62, +24.25 &                        (4.08, 30.8, 3.13) &                              0.67 \\
122469:31925079 &  164.18, +6.32 & 56443.56 &                               0.00 &                               0.00 &                  164.74, +6.58 &                         (3.06, 3.9, 2.05) &                              0.38 \\
122604:17469985 &  93.74, +14.17 & 56470.11 &                               0.00 &                               0.00 &                  93.60, +14.17 &                        (2.90, 25.4, 2.94) &                              0.65 \\
122605:60656774 &  155.35, +3.73 & 56470.43 &                               0.00 &                               0.00 &                  156.14, +3.44 &                        (4.56, 41.5, 4.48) &                              0.63 \\
122663:58987459 &   77.87, -2.43 & 56484.53 &                               0.00 &                               0.00 &                   78.64, -2.28 &                        (2.52, 29.5, 3.23) &                              0.33 \\
122768:38570780 &  122.87, +6.32 & 56504.07 &                               0.00 &                               0.35 &                  124.17, +6.44 &                        (6.85, 58.0, 3.55) &                              0.68 \\
122772:38615684 &  214.98, +7.75 & 56505.26 &                               0.00 &                               0.00 &                  213.99, +7.72 &                        (6.25, 51.8, 3.72) &                              0.88 \\
122793:12516001 & 129.02, +13.36 & 56508.81 &                               0.00 &                               0.00 &                 128.31, +12.70 &                         (4.69, 1.6, 1.41) &                              0.82 \\
122818:16944987 &   26.59, +9.22 & 56512.34 &                               0.00 &                               0.00 &                   26.43, +9.95 &                         (1.82, 8.4, 2.08) &                              0.55 \\
122895:69011827 &   91.32, +0.56 & 56526.41 &                               0.00 &                               0.00 &                             -- &                            (0.00, --, --) &                              0.00 \\
 122973:6578595 & 130.17, -10.54 & 56542.79 &                               0.00 &                               0.00 &                             -- &                            (0.00, --, --) &                              0.00 \\
 123107:8144529 &  32.92, +10.28 & 56579.91 &                               0.00 &                               0.00 &                             -- &                            (0.00, --, --) &                              0.00 \\
123145:34573112 & 301.90, +11.61 & 56588.56 &                               0.00 &                               0.00 &                 301.99, +11.42 &                         (0.38, 9.3, 3.20) &                              0.19 \\
123228:40504010 & 342.73, +41.81 & 56604.55 &                               0.00 &                               0.00 &                             -- &                            (0.00, --, --) &                              0.00 \\
123240:49730417 & 129.24, -17.27 & 56608.03 &                               0.00 &                               0.00 &                             -- &                            (0.00, --, --) &                              0.00 \\
123281:52248792 & 285.16, +19.47 & 56620.15 &                               0.00 &                               0.00 &                             -- &                            (0.00, --, --) &                              0.00 \\
123334:75182550 & 288.98, -14.21 & 56630.47 &                               0.00 &                               1.15 &                 288.84, -14.15 &                        (4.67, 10.2, 2.61) &                              0.59 \\
123620:47444787 &  192.26, -2.69 & 56658.40 &                               0.00 &                               0.00 &                  190.98, -2.58 &                        (6.28, 42.9, 4.83) &                              1.11 \\
123662:27529428 &  37.90, +78.97 & 56660.89 &                               0.00 &                               0.44 &                  22.50, +82.59 &                         (5.61, 3.5, 1.35) &                              0.54 \\
123751:34239163 &  344.66, +1.57 & 56665.31 &                               0.00 &                               0.68 &                             -- &                            (0.00, --, --) &                              0.00 \\
123762:72626160 & 293.12, +33.02 & 56666.50 &                               0.00 &                               0.00 &                             -- &                            (0.00, --, --) &                              0.00 \\
123867:11659459 &  337.59, +0.71 & 56671.88 &                               0.00 &                               0.00 &                             -- &                            (0.00, --, --) &                              0.00 \\
123986:63557286 & 138.82, +37.45 & 56679.15 &                               0.00 &                               0.00 &                 140.19, +37.54 &                        (2.78, 19.7, 2.92) &                              0.08 \\
123986:77999595 & 220.29, -86.07 & 56679.20 &                               0.00 &                               0.00 &                 216.74, -85.98 &                         (1.01, 3.2, 2.88) &                              0.22 \\
124136:15174527 & 349.58, -13.55 & 56691.79 &                               0.00 &                               0.00 &                 350.71, -15.13 &                        (9.33, 30.3, 4.18) &                              1.00 \\
124221:76481548 & 202.59, +13.06 & 56701.81 &                               0.00 &                               0.00 &                 204.29, +13.07 &                        (1.88, 27.6, 3.74) &                              0.07 \\
124296:49827757 & 118.83, +32.58 & 56711.92 &                               0.00 &                               0.01 &                          Excl. &                                     Excl. &                             Excl. \\
124340:19400842 & 308.06, +32.93 & 56723.92 &                               0.00 &                               0.00 &                 308.31, +31.63 &                        (4.38, 35.0, 3.25) &                              0.33 \\
124463:35466777 & 225.70, +51.06 & 56740.09 &                               0.00 &                               0.69 &                 227.23, +50.73 &                        (4.10, 24.3, 2.84) &                              0.21 \\
124547:66087371 &   2.11, +81.22 & 56757.10 &                               0.00 &                               0.18 &                          Excl. &                                     Excl. &                             Excl. \\
124569:22040903 & 146.95, +15.91 & 56758.57 &                               0.00 &                               0.51 &                 149.20, +16.66 &                        (3.55, 47.6, 3.69) &                              0.22 \\
 124643:5182926 &   6.28, +16.57 & 56767.86 &                               0.00 &                               0.00 &                   6.47, +14.77 &                        (9.66, 58.1, 3.02) &                              0.80 \\
124693:33063779 & 162.30, +46.57 & 56780.96 &                               0.00 &                               0.00 &                 161.65, +42.97 &                       (13.56, 65.4, 3.92) &                              1.50 \\
124829:42577032 &    9.71, +7.56 & 56811.14 &                               0.00 &                               0.00 &                    9.69, +7.60 &                         (5.98, 1.9, 1.67) &                              1.41 \\
124852:74171655 &  106.26, +1.31 & 56817.64 &                               0.00 &                               0.00 &                  105.77, +1.19 &                        (4.77, 34.5, 3.04) &                              0.60 \\
124861:32863663 & 110.65, +11.45 & 56819.20 &                               0.00 &                               0.00 &                             -- &                            (0.00, --, --) &                              0.00 \\
124994:15096469 & 157.07, +53.62 & 56842.30 &                               0.00 &                               0.00 &                 155.88, +53.89 &                         (1.07, 5.2, 2.41) &                              0.02 \\
125001:73159253 &   25.88, +2.54 & 56843.67 &                               0.00 &                               0.00 &                   26.26, +1.08 &                       (11.07, 67.5, 3.74) &                              1.74 \\
125011:43302321 & 240.86, +14.17 & 56845.50 &                               0.00 &                               0.00 &                 240.45, +14.70 &                        (5.79, 44.6, 3.52) &                              0.86 \\
125046:64570437 &   0.79, +15.60 & 56851.56 &                               0.00 &                               0.00 &                             -- &                            (0.00, --, --) &                              0.00 \\
125071:31397276 & 101.82, -32.89 & 56859.76 &                               0.00 &                               0.00 &                 100.46, -37.25 &                       (11.95, 19.7, 3.22) &                              0.46 \\
125205:19617837 &  271.45, +1.87 & 56889.38 &                               0.00 &                               0.00 &                  271.16, +1.95 &                        (3.68, 33.9, 3.55) &                              0.35 \\
125338:65606123 &  169.72, -1.60 & 56923.72 &                               0.00 &                               0.00 &                             -- &                            (0.00, --, --) &                              0.00 \\
125349:12316485 &   50.89, -0.63 & 56927.16 &                               0.00 &                               0.00 &                   50.77, -0.47 &                        (7.95, 49.0, 3.67) &                              0.89 \\
125422:26630854 &   63.85, +3.21 & 56942.75 &                               0.00 &                               0.00 &                   65.06, +2.70 &                        (5.22, 52.3, 3.24) &                              0.93 \\
125544:54236696 &  253.43, +6.43 & 56971.30 &                              Excl. &                               0.00 &                             -- &                            (0.00, --, --) &                              0.00 \\
125558:19184163 & 221.48, +28.00 & 56975.26 &                               0.00 &                               0.83 &                 221.97, +27.36 &                        (1.18, 18.1, 3.82) &                              0.07 \\
125693:21174828 & 246.36, +17.23 & 56999.67 &                               0.00 &                               0.00 &                 246.79, +17.77 &                        (4.65, 40.3, 3.46) &                              0.75 \\
 125709:6405911 &  318.12, +1.57 & 57001.85 &                              Excl. &                               0.00 &                  317.56, +2.06 &                         (1.50, 4.1, 2.03) &                              0.12 \\
125757:12938244 &  179.08, -1.94 & 57012.41 &                               0.00 &                               0.80 &                  178.71, -2.20 &                        (1.05, 20.1, 3.52) &                              0.31 \\
125796:16042897 &  318.74, +2.91 & 57024.80 &                               0.00 &                               2.17 &                  319.75, +2.99 &                        (0.12, 13.5, 3.06) &                              0.06 \\
125800:83097666 & 272.11, +28.76 & 57026.40 &                               0.00 &                               0.00 &                 272.00, +28.98 &                        (1.22, 16.8, 4.33) &                              0.20 \\
125929:11025256 &  152.53, +4.33 & 57040.51 &                               0.00 &                               0.95 &                  153.48, +4.10 &                         (0.27, 4.7, 2.21) &                              0.10 \\
125934:55750717 &  286.92, +6.43 & 57041.37 &                               0.00 &                               0.00 &                             -- &                            (0.00, --, --) &                              0.00 \\
 125940:7449833 &  95.89, +14.13 & 57042.98 &                               0.00 &                               0.00 &                             -- &                            (0.00, --, --) &                              0.00 \\
125968:61389842 &  100.37, +4.59 & 57049.48 &                               0.00 &                               0.00 &                   99.17, +4.71 &                        (0.25, 28.3, 3.80) &                              0.15 \\
125973:80686964 &  358.51, +6.39 & 57051.23 &                               0.00 &                               2.10 &                  358.16, +6.20 &                         (3.72, 4.9, 2.07) &                              0.07 \\
126082:40384770 & 237.75, +55.11 & 57078.00 &                               0.00 &                               0.00 &                 237.01, +57.02 &                        (2.94, 32.9, 3.74) &                              0.10 \\
126148:74589372 &  127.05, -3.36 & 57094.32 &                               0.00 &                               0.00 &                  126.23, -3.12 &                        (0.39, 22.3, 4.16) &                              0.29 \\
126308:20883844 &  31.07, +15.02 & 57140.59 &                               0.00 &                               1.19 &                  34.56, +14.76 &                        (6.47, 55.9, 4.01) &                              0.57 \\
126370:61611641 &  91.49, +12.14 & 57157.94 &                               0.00 &                               0.00 &                  91.92, +12.20 &                        (0.66, 20.3, 3.40) &                              0.28 \\
126405:50771014 &  139.79, -1.49 & 57168.02 &                               0.00 &                               0.00 &                  139.57, -1.19 &                        (0.35, 15.7, 4.83) &                              0.05 \\
126423:50123696 &  333.37, +9.63 & 57174.02 &                               0.00 &                               0.62 &                  334.68, +9.90 &                        (5.61, 24.0, 2.68) &                              1.04 \\
 126456:1581608 &   49.53, +0.30 & 57182.03 &                               0.00 &                               0.00 &                   50.38, +0.01 &                        (4.81, 46.2, 3.15) &                              0.98 \\
126456:35956042 &  245.43, +0.22 & 57182.18 &                               0.00 &                               1.64 &                             -- &                            (0.00, --, --) &                              0.00 \\
126514:73786337 &   71.89, +0.86 & 57198.64 &                               0.00 &                               0.93 &                   73.17, +0.33 &                        (8.22, 50.7, 3.95) &                              0.79 \\
126515:20091930 & 306.43, +19.08 & 57198.73 &                               0.00 &                               0.00 &                 306.41, +18.57 &                        (3.08, 34.7, 3.21) &                              0.47 \\
126620:19175993 & 326.29, +26.36 & 57217.91 &                               0.00 &                               0.00 &                 325.90, +26.00 &                        (2.71, 25.6, 4.65) &                              0.40 \\
126703:23477554 & 221.75, -17.15 & 57243.32 &                               0.00 &                               0.00 &                             -- &                            (0.00, --, --) &                              0.00 \\
126717:29833358 & 317.59, +30.09 & 57246.32 &                               0.00 &                               0.42 &                 320.48, +30.77 &                        (1.87, 26.7, 3.33) &                              0.07 \\
126718:53509959 &  328.27, +6.17 & 57246.76 &                               0.00 &                               1.45 &                             -- &                            (0.00, --, --) &                              0.00 \\
126769:62990844 &  325.90, -2.35 & 57257.62 &                               0.00 &                               1.25 &                  327.82, -0.59 &                        (5.58, 55.8, 3.42) &                              0.47 \\
126798:67205547 &  54.76, +34.00 & 57265.22 &                               0.00 &                               0.00 &                  54.86, +34.60 &                        (7.15, 51.7, 3.85) &                              1.54 \\
126812:38566267 & 133.77, +28.08 & 57269.76 &                               0.00 &                               0.00 &                 133.92, +27.95 &                        (0.58, 16.5, 4.18) &                              0.38 \\
 126848:8902133 & 129.68, +30.35 & 57279.87 &                               0.00 &                               0.00 &                 129.81, +30.62 &                        (2.50, 28.9, 4.08) &                              0.31 \\
126860:17477746 &   49.83, -2.95 & 57283.55 &                               0.00 &                               0.87 &                   50.57, -2.86 &                        (1.02, 19.2, 3.63) &                              0.12 \\
126863:14025085 & 279.54, +30.35 & 57284.21 &                               0.00 &                               0.00 &                             -- &                            (0.00, --, --) &                              0.00 \\
126878:30560694 &  103.23, +3.96 & 57288.03 &                               0.00 &                               0.00 &                             -- &                            (0.00, --, --) &                              0.00 \\
 126911:6416344 &  194.55, -4.56 & 57291.90 &                               0.00 &                               0.00 &                             -- &                            (0.00, --, --) &                              0.00 \\
 126976:9961785 & 178.72, +52.37 & 57308.12 &                               0.00 &                               0.00 &                             -- &                            (0.00, --, --) &                              0.00 \\
126989:59479399 & 197.53, +19.95 & 57312.68 &                               0.00 &                               1.98 &                 197.30, +19.13 &                        (4.71, 20.2, 2.65) &                              0.47 \\
127112:69211097 &  76.16, +12.71 & 57340.87 &                               0.00 &                               0.00 &                  75.24, +12.64 &                        (9.49, 23.5, 2.32) &                              2.15 \\
 127154:9907321 &  262.05, -2.24 & 57348.53 &                               0.00 &                               0.00 &                  262.34, -2.43 &                        (0.59, 19.4, 3.65) &                              0.27 \\
127357:17650073 &   79.41, +5.00 & 57391.44 &                               0.00 &                               0.00 &                             -- &                            (0.00, --, --) &                              0.00 \\
127495:54505431 & 263.76, -14.90 & 57415.18 &                               0.00 &                               0.00 &                 263.50, -14.79 &                         (0.72, 6.8, 3.83) &                              0.03 \\
127603:48070937 & 311.87, +60.06 & 57443.88 &                               0.00 &                               0.00 &                             -- &                            (0.00, --, --) &                              0.00 \\
127650:18431575 &  91.32, +10.47 & 57454.70 &                               0.00 &                               0.00 &                   92.83, +8.60 &                        (6.76, 56.0, 3.63) &                              0.32 \\
127742:55820225 & 151.22, +15.48 & 57478.57 &                               0.00 &                               0.00 &                             -- &                            (0.00, --, --) &                              0.00 \\
127790:45902607 &  235.63, -4.07 & 57488.73 &                               0.00 &                               0.00 &                  234.98, -4.43 &                        (1.14, 10.5, 2.37) &                              0.56 \\
127853:67093193 &  240.29, +9.71 & 57505.24 &                               0.00 &                               0.00 &                             -- &                            (0.00, --, --) &                              0.00 \\
127910:25056152 &  352.88, +1.90 & 57518.66 &                               0.00 &                               0.00 &                  354.13, +1.61 &                        (0.87, 31.8, 3.85) &                              0.19 \\
128034:69069846 &   16.52, +4.67 & 57551.43 &                               0.00 &                               0.00 &                             -- &                            (0.00, --, --) &                              0.00 \\
128065:17929326 & 214.76, +40.82 & 57553.53 &                               0.00 &                               0.00 &                 216.63, +39.31 &                         (2.73, 8.4, 2.40) &                              0.07 \\
128067:65335330 & 304.32, +12.64 & 57554.40 &                               0.00 &                               0.00 &                 304.74, +11.72 &                        (0.30, 30.1, 3.75) &                              0.12 \\
128209:54386016 &  351.43, +0.60 & 57576.17 &                               0.00 &                               0.00 &                  351.27, +0.24 &                        (0.71, 17.8, 3.90) &                              0.12 \\
128253:44251618 &  60.25, +29.23 & 57589.91 &                               0.00 &                               0.17 &                          Excl. &                                     Excl. &                             Excl. \\
128278:49218472 & 113.12, +14.67 & 57596.34 &                               0.00 &                               1.51 &                             -- &                            (0.00, --, --) &                              0.00 \\
 128290:6888376 &  214.58, -0.30 & 57600.08 &                               0.00 &                               0.00 &                             -- &                            (0.00, --, --) &                              0.00 \\
128292:15195696 & 312.63, +20.07 & 57600.78 &                               0.00 &                               0.66 &                 311.82, +18.76 &                         (2.63, 3.8, 1.97) &                              0.13 \\
128311:26552458 &  122.78, -0.71 & 57606.51 &                               0.00 &                               0.00 &                             -- &                            (0.00, --, --) &                              0.00 \\
 128334:9739548 &  86.99, +48.83 & 57612.68 &                               0.00 &                               0.97 &                          Excl. &                                     Excl. &                             Excl. \\
128340:58537957 & 200.04, -32.13 & 57614.91 &                               0.00 &                               0.79 &                 202.31, -31.47 &                         (1.68, 9.7, 4.33) &                              0.01 \\
128547:14557367 &  241.13, +1.34 & 57655.74 &                               0.00 &                               0.73 &                  242.25, +2.29 &                        (3.50, 31.9, 2.93) &                              0.05 \\
128567:21044380 & 192.57, +37.12 & 57662.44 &                               0.00 &                               0.00 &                 191.83, +35.09 &                        (1.67, 35.8, 4.03) &                              0.18 \\
128606:54200591 &  190.06, -7.48 & 57673.61 &                               0.00 &                               0.24 &                  190.38, -7.73 &                        (6.66, 16.9, 2.49) &                              0.50 \\
128632:62194858 & 121.42, +23.72 & 57682.31 &                               0.00 &                               1.06 &                 122.56, +22.21 &                        (8.65, 12.7, 2.11) &                              1.22 \\
128651:44166050 &  119.00, +1.53 & 57688.57 &                               0.00 &                               1.53 &                  116.88, +0.44 &                        (5.28, 53.3, 3.04) &                              0.49 \\
128672:38561326 &  40.87, +12.52 & 57695.38 &                               0.00 &                               0.00 &                  41.26, +12.52 &                         (6.30, 5.7, 1.59) &                              1.34 \\
128755:32356079 &   78.66, +1.60 & 57709.33 &                               0.00 &                               0.00 &                   79.44, +2.69 &                        (9.63, 27.2, 2.58) &                              1.36 \\
128785:76992952 &  140.01, -0.11 & 57717.43 &                               0.00 &                               0.00 &                  139.39, -0.30 &                        (2.63, 37.8, 4.22) &                              0.43 \\
128796:26367207 & 257.55, +73.27 & 57719.66 &                               0.00 &                               0.01 &                          Excl. &                                     Excl. &                             Excl. \\
128906:80127519 &  46.36, +15.25 & 57732.84 &                               0.00 &                               0.97 &                  46.93, +16.10 &                        (3.90, 48.1, 3.16) &                              0.64 \\
128967:17750653 &  61.79, +17.78 & 57746.54 &                               0.00 &                               0.00 &                  62.20, +17.27 &                        (1.90, 22.6, 3.07) &                              0.19 \\
129020:20626582 &  309.95, +8.16 & 57758.14 &                               0.00 &                               0.21 &                  314.47, +7.48 &                        (4.67, 58.7, 3.85) &                              0.21 \\
129144:66284903 & 180.35, +33.20 & 57790.55 &                               0.00 &                               0.00 &                 178.95, +31.89 &                       (16.85, 82.0, 3.89) &                              2.76 \\
129153:11436007 &  99.67, +16.84 & 57792.13 &                               0.00 &                               0.00 &                 101.01, +17.10 &                        (0.84, 15.9, 2.70) &                              0.06 \\
129154:45298080 &   92.81, +4.59 & 57792.60 &                               0.00 &                               0.00 &                   92.24, +4.69 &                        (0.41, 19.1, 2.92) &                              0.22 \\
129232:51118868 &  205.09, +4.26 & 57811.06 &                               0.00 &                               0.00 &                  205.26, +4.08 &                       (10.80, 68.2, 4.10) &                              1.96 \\
129267:34234803 &  155.35, +5.53 & 57820.92 &                               0.00 &                               1.73 &                  155.02, +5.38 &                        (5.14, 39.6, 3.08) &                              0.87 \\
129307:80305071 &  98.26, -15.06 & 57833.31 &                               0.00 &                               0.00 &                  99.09, -14.99 &                        (0.62, 10.7, 3.84) &                              0.03 \\
129420:59928529 &  240.95, +5.53 & 57865.65 &                               0.00 &                               0.00 &                  243.42, +5.70 &                        (4.47, 46.4, 3.72) &                              0.43 \\
129434:58903823 &    5.32, -0.60 & 57870.31 &                               0.00 &                               0.00 &                    3.19, -0.37 &                        (4.54, 37.8, 3.99) &                              0.33 \\
129506:21161807 & 311.97, +18.60 & 57887.17 &                               0.00 &                               0.00 &                 312.94, +20.21 &                         (2.34, 5.2, 2.10) &                              0.27 \\
129506:49650572 & 227.37, +30.65 & 57887.30 &                               0.00 &                               0.00 &                 227.11, +30.35 &                        (1.11, 19.3, 3.73) &                              0.37 \\
129550:51402681 & 178.59, +26.49 & 57900.07 &                               0.00 &                               0.00 &                 177.80, +23.37 &                        (9.22, 36.1, 2.57) &                              1.14 \\
 129654:8332254 &  74.97, +25.08 & 57925.19 &                               0.00 &                               0.30 &                  75.89, +27.51 &                        (5.19, 11.3, 2.24) &                              0.07 \\
129677:55886338 &  280.99, +8.80 & 57930.52 &                               0.00 &                               0.00 &                  281.80, +8.93 &                        (4.05, 41.1, 3.35) &                              0.77 \\
129701:49353375 & 230.45, +23.36 & 57938.29 &                               0.00 &                               0.00 &                 230.10, +23.64 &                         (0.70, 7.6, 2.49) &                              0.18 \\
129777:67372962 & 208.39, +25.16 & 57951.82 &                               0.00 &                               1.62 &                 208.29, +25.68 &                         (4.20, 4.2, 1.98) &                              0.73 \\
129855:33565191 &    1.10, +4.63 & 57968.08 &                               0.00 &                               0.72 &                    3.69, +4.48 &                        (1.29, 24.6, 3.34) &                              0.12 \\
129878:40814378 &   21.27, -2.28 & 57974.60 &                               0.00 &                               0.00 &                   21.42, -2.22 &                        (3.34, 30.5, 3.52) &                              0.81 \\
129915:72252401 &  26.98, +18.88 & 57984.28 &                               0.00 &                               0.77 &                  27.56, +18.33 &                        (7.11, 51.9, 3.37) &                              1.06 \\
129933:32926212 &  41.92, +12.37 & 57989.55 &                               0.00 &                               0.00 &                  40.96, +12.49 &                        (12.56, 4.8, 1.43) &                              2.06 \\
130033:50579430 &   77.43, +5.79 & 58018.87 &                               0.00 &                               0.00 &                   77.52, +5.68 &                        (1.63, 17.5, 2.75) &                              0.44 \\
 130034:7858514 &  173.45, -2.54 & 58019.02 &                               0.00 &                               0.00 &                  172.34, -2.37 &                         (1.10, 6.0, 2.28) &                              0.16 \\
130092:30964247 & 132.63, +17.23 & 58032.31 &                               0.00 &                               0.00 &                 130.81, +16.78 &                        (3.97, 59.1, 3.93) &                              0.81 \\
130126:56068624 & 162.91, -15.48 & 58041.07 &                               0.00 &                               0.00 &                 164.15, -17.19 &                        (5.28, 20.0, 3.24) &                              0.24 \\
130172:52824390 &  294.52, +2.05 & 58054.76 &                               0.00 &                               1.87 &                  292.80, +1.44 &                       (11.38, 64.8, 3.62) &                              1.53 \\
130214:17569642 &  340.14, +7.44 & 58063.78 &                               0.00 &                               0.96 &                             -- &                            (0.00, --, --) &                              0.00 \\
130220:11599241 & 269.65, -20.70 & 58065.75 &                               0.00 &                               0.00 &                             -- &                            (0.00, --, --) &                              0.00 \\
130561:84363835 &  206.10, +3.92 & 58135.75 &                               0.00 &                               0.00 &                  205.45, +4.05 &                        (8.52, 62.8, 4.09) &                              1.81 \\
130588:44934051 &   77.12, +8.01 & 58141.68 &                               0.00 &                               0.00 &                   76.76, +8.02 &                        (1.40, 21.5, 3.81) &                              0.35 \\
130597:55489144 & 207.51, +23.77 & 58143.98 &                               0.00 &                               0.00 &                 208.12, +23.35 &                        (7.11, 57.0, 3.35) &                              1.62 \\
130639:37366501 &  17.40, -10.54 & 58154.00 &                               0.00 &                               0.00 &                  17.80, -10.61 &                        (2.27, 12.1, 3.65) &                              0.22 \\
130684:80612787 &   66.97, +6.09 & 58162.38 &                               0.00 &                               0.00 &                   66.86, +6.29 &                        (3.75, 33.5, 3.19) &                              0.37 \\
130743:25740057 & 294.79, +26.40 & 58177.57 &                               0.00 &                               0.00 &                             -- &                            (0.00, --, --) &                              0.00 \\
130797:67186364 &  287.18, +5.53 & 58190.68 &                               0.00 &                               0.00 &                             -- &                            (0.00, --, --) &                              0.00 \\
130801:16289732 &   58.71, +0.78 & 58191.80 &                               0.00 &                               0.60 &                   59.48, +0.90 &                        (1.58, 29.2, 3.45) &                              0.18 \\
 130807:3160265 &  271.71, -1.42 & 58193.24 &                               0.00 &                               1.43 &                  270.93, -1.64 &                        (1.78, 18.4, 2.81) &                              0.16 \\
130912:76035104 &  218.50, +0.56 & 58218.78 &                               0.00 &                               0.00 &                  216.80, +1.02 &                        (5.18, 56.3, 3.09) &                              1.02 \\
130932:35022693 &  305.73, -4.41 & 58225.28 &                               0.00 &                               0.56 &                  305.21, -4.81 &                        (3.67, 34.5, 4.23) &                              0.77 \\
131096:32665194 &  312.19, +0.30 & 58266.51 &                              Excl. &                              Excl. &                          Excl. &                                     Excl. &                             Excl. \\
131134:60192271 &   69.08, -1.08 & 58277.60 &                               0.00 &                               0.00 &                             -- &                            (0.00, --, --) &                              0.00 \\
131145:43542963 &  338.69, +3.73 & 58281.19 &                               0.00 &                               0.00 &                  342.41, +2.95 &                        (6.19, 51.0, 3.05) &                              0.29 \\
 131165:9342044 &  38.06, +11.53 & 58282.98 &                               0.00 &                               0.26 &                  40.99, +12.33 &                        (14.93, 6.7, 1.48) &                              1.98 \\
131321:73241305 &  74.14, -17.74 & 58327.84 &                               0.00 &                               0.00 &                             -- &                            (0.00, --, --) &                              0.00 \\
131360:57649537 & 100.37, +11.15 & 58337.20 &                               0.00 &                               0.54 &                  99.40, +10.76 &                         (7.58, 8.8, 2.13) &                              0.76 \\
131475:34507973 &  144.98, -2.39 & 58369.83 &                               0.00 &                               1.31 &                  145.34, -2.92 &                        (0.26, 20.2, 3.33) &                              0.08 \\
131477:57977901 & 141.37, +26.94 & 58370.60 &                               0.00 &                               0.00 &                             -- &                            (0.00, --, --) &                              0.00 \\
131519:27561842 & 258.40, +32.84 & 58380.07 &                               0.00 &                               0.00 &                 258.75, +33.09 &                        (2.57, 28.3, 3.98) &                              0.58 \\
131602:39194539 &   77.08, +1.23 & 58399.78 &                               0.00 &                               0.00 &                   77.45, +2.47 &                        (5.32, 43.6, 2.86) &                              0.54 \\
131624:12296708 & 225.22, -34.95 & 58405.49 &                               0.00 &                               0.00 &                 225.29, -35.40 &                        (3.21, 11.1, 2.51) &                              0.27 \\
131653:53411354 &  270.18, -8.42 & 58414.69 &                               0.00 &                               0.00 &                  268.82, -8.10 &                        (1.59, 25.5, 3.33) &                              0.34 \\
131653:63430929 &  78.27, +21.54 & 58414.74 &                               0.00 &                               0.00 &                             -- &                            (0.00, --, --) &                              0.00 \\
131746:47815348 &   6.02, +18.84 & 58436.94 &                               0.00 &                               0.00 &                   4.87, +19.22 &                        (2.00, 39.5, 3.62) &                              0.40 \\
131764:48222919 &  25.71, +11.72 & 58442.71 &                               0.00 &                               0.00 &                  26.43, +10.07 &                         (4.29, 8.4, 2.10) &                              0.16 \\
 131767:7377802 & 324.58, +51.74 & 58442.94 &                               0.00 &                               0.42 &                 321.11, +51.40 &                        (7.27, 32.3, 2.74) &                              0.29 \\
131768:80926380 & 132.19, +32.93 & 58443.58 &                               0.00 &                               0.74 &                 129.38, +33.15 &                        (4.30, 37.7, 4.02) &                              0.08 \\
131913:38717715 & 316.41, -31.00 & 58464.09 &                               0.00 &                               0.77 &                 315.19, -31.45 &                        (7.07, 19.9, 2.79) &                              0.69 \\
132043:80110393 &   56.91, -0.82 & 58496.09 &                               0.00 &                               0.00 &                   56.95, -1.00 &                        (2.13, 26.7, 3.24) &                              0.39 \\
 132077:9759013 & 307.44, -32.22 & 58507.16 &                               0.00 &                               1.05 &                 308.19, -32.10 &                        (1.90, 10.7, 2.25) &                              0.25 \\
132128:69441226 & 245.08, +38.78 & 58515.02 &                               0.00 &                               0.00 &                 244.72, +38.19 &                         (0.80, 4.2, 2.07) &                              0.43 \\
132206:36575563 &  228.25, -4.14 & 58528.67 &                               0.00 &                               0.00 &                  227.99, -4.18 &                        (0.35, 16.9, 3.65) &                              0.74 \\
132229:66688965 & 268.59, -17.00 & 58535.35 &                               0.00 &                               1.18 &                 267.70, -16.54 &                        (4.86, 10.6, 2.62) &                              0.55 \\
132237:23969090 & 155.21, +19.67 & 58537.85 &                               0.00 &                               0.00 &                             -- &                            (0.00, --, --) &                              0.00 \\
132321:28454069 &   81.25, +3.21 & 58559.83 &                               0.00 &                               0.66 &                   79.45, +2.71 &                       (11.55, 22.7, 2.53) &                              1.22 \\
132427:70353420 & 310.61, +12.22 & 58583.44 &                               0.00 &                               0.00 &                 310.89, +11.20 &                        (5.52, 37.6, 2.82) &                              0.48 \\
132437:16335312 & 219.33, +11.72 & 58586.45 &                               0.00 &                               0.69 &                             -- &                            (0.00, --, --) &                              0.00 \\
132437:67132865 & 245.57, +21.98 & 58586.66 &                               0.00 &                               0.68 &                 245.45, +22.11 &                        (0.41, 13.5, 4.05) &                              0.17 \\
132443:12627143 &  154.86, +5.27 & 58588.44 &                               0.00 &                               0.00 &                  153.79, +5.25 &                        (8.56, 56.5, 3.71) &                              0.97 \\
 132465:3856549 & 166.90, +17.39 & 58595.25 &                               0.00 &                               0.00 &                 169.03, +17.05 &                        (4.03, 21.6, 2.59) &                              0.30 \\
132508:42419327 &  120.19, +6.43 & 58606.72 &                               0.00 &                               0.00 &                  120.24, +6.12 &                        (1.41, 23.6, 2.96) &                              0.51 \\
  132518:766165 &  65.17, -37.26 & 58607.77 &                               0.00 &                               0.00 &                  65.79, -36.90 &                        (2.21, 10.0, 3.16) &                              0.30 \\
132577:42662743 & 127.88, +12.60 & 58618.45 &                               0.00 &                               0.00 &                             -- &                            (0.00, --, --) &                              0.00 \\
 132684:5635104 & 312.19, +26.57 & 58647.83 &                               0.00 &                               0.98 &                             -- &                            (0.00, --, --) &                              0.00 \\
132707:54984442 & 343.52, +10.28 & 58653.55 &                               0.00 &                               0.61 &                  344.37, +9.51 &                        (3.07, 35.5, 4.13) &                              0.20 \\
 132768:5390846 &  29.12, +84.56 & 58663.81 &                               0.00 &                               0.00 &                  22.43, +82.58 &                         (5.46, 2.9, 1.37) &                              0.89 \\
132792:60166398 & 161.81, +26.90 & 58668.78 &                               0.00 &                               0.00 &                 163.39, +27.89 &                        (5.99, 29.7, 2.61) &                              0.74 \\
132814:44222682 &  76.64, +12.75 & 58676.05 &                               0.00 &                               0.00 &                  75.35, +12.61 &                       (12.60, 18.2, 2.07) &                              1.40 \\
132910:57145925 & 226.14, +10.77 & 58694.87 &                               0.00 &                               0.00 &                 225.60, +10.76 &                        (0.18, 14.9, 4.05) &                              0.11 \\
132974:67924813 &  148.54, +1.45 & 58714.73 &                               0.00 &                               0.00 &                             -- &                            (0.00, --, --) &                              0.00 \\
   133091:81419 & 167.30, -22.27 & 58748.40 &                               0.00 &                               0.39 &                 167.88, -24.59 &                        (6.04, 16.7, 3.43) &                              0.19 \\
133092:52499868 &    5.71, -1.53 & 58748.96 &                               0.00 &                               0.00 &                             -- &                            (0.00, --, --) &                              0.00 \\
133119:22683750 & 313.99, +12.79 & 58757.84 &                               0.00 &                               1.12 &                 317.86, +13.21 &                         (8.21, 6.1, 1.95) &                              0.95 \\
133331:47828126 &  229.31, +3.77 & 58806.04 &                               0.00 &                               0.32 &                  233.77, +2.79 &                        (1.75, 48.0, 3.33) &                              0.01 \\
133348:80807014 &   27.03, +0.07 & 58809.95 &                               0.00 &                               2.08 &                   26.27, +0.92 &                       (11.42, 69.0, 3.75) &                              1.83 \\
133394:27261780 &   80.16, +2.87 & 58821.95 &                               0.00 &                               0.00 &                   79.48, +2.68 &                        (4.77, 27.5, 2.83) &                              0.76 \\
133433:29047901 & 286.83, +58.45 & 58832.47 &                               0.00 &                               1.94 &                 288.60, +57.91 &                        (4.99, 27.3, 2.75) &                              0.80 \\
133572:82361476 &  48.47, +20.11 & 58848.46 &                               0.00 &                               0.85 &                  45.53, +16.96 &                        (4.80, 44.3, 3.05) &                              0.10 \\
133609:37927131 & 165.45, +11.80 & 58857.99 &                               0.00 &                               0.00 &                 166.96, +11.27 &                        (2.94, 32.9, 3.72) &                              0.25 \\
 133634:1410505 & 116.02, +29.18 & 58865.46 &                               0.00 &                               0.80 &                 116.27, +29.36 &                        (0.35, 16.9, 4.00) &                              0.27 \\
133644:43767651 &  67.41, -14.59 & 58868.78 &                               0.00 &                               0.00 &                             -- &                            (0.00, --, --) &                              0.00 \\
133945:24635982 & 242.58, +11.61 & 58949.97 &                               0.00 &                               0.10 &                          Excl. &                                     Excl. &                             Excl. \\
133985:60770138 &   87.93, +8.23 & 58960.02 &                               0.00 &                               0.00 &                   89.53, +8.04 &                        (6.94, 57.5, 3.55) &                              1.00 \\
134013:16038252 &  99.97, +53.72 & 58964.98 &                               0.00 &                               1.09 &                             -- &                            (0.00, --, --) &                              0.00 \\
134081:58268464 & 295.18, +15.79 & 58981.31 &                               0.00 &                               0.00 &                             -- &                            (0.00, --, --) &                              0.00 \\
134116:58596690 &  338.64, +1.75 & 58992.10 &                               0.00 &                               0.46 &                  347.73, +1.25 &                        (7.11, 54.1, 3.09) &                              0.36 \\
134139:35473338 & 255.37, +26.61 & 58999.33 &                               0.00 &                               0.00 &                 255.89, +26.68 &                        (5.78, 35.7, 3.17) &                              0.57 \\
134187:72386329 &  33.84, +31.61 & 59014.53 &                               0.00 &                               0.00 &                  31.22, +32.39 &                       (14.57, 75.1, 3.63) &                              2.40 \\
134191:17593623 &  142.95, +3.66 & 59015.62 &                               0.00 &                               0.00 &                  143.08, +3.53 &                        (5.79, 26.8, 2.72) &                              1.08 \\
134207:33533447 & 162.11, +11.95 & 59020.13 &                               0.00 &                               0.00 &                 162.09, +12.02 &                        (2.93, 34.5, 3.75) &                              0.79 \\
134354:59221243 & 157.25, +47.75 & 59067.58 &                               0.00 &                               0.82 &                             -- &                            (0.00, --, --) &                              0.00 \\
134482:27754576 &  51.11, +38.11 & 59103.60 &                               0.00 &                               0.00 &                  50.67, +39.59 &                        (4.99, 26.9, 2.94) &                              0.52 \\
134498:12605830 & 109.78, +14.36 & 59108.86 &                               0.00 &                               0.00 &                             -- &                            (0.00, --, --) &                              0.00 \\
134512:71996695 & 195.29, +26.24 & 59113.80 &                               0.00 &                               0.00 &                 194.19, +25.77 &                        (1.14, 13.7, 2.47) &                              0.17 \\
134533:53384881 &   96.46, -4.33 & 59118.33 &                               2.17 &                               0.00 &                             -- &                            (0.00, --, --) &                              0.00 \\
134535:41069485 & 184.75, +32.93 & 59118.94 &                               0.00 &                               0.00 &                 184.22, +32.35 &                         (1.29, 5.5, 2.22) &                              0.17 \\
134552:68615710 &   29.53, +3.47 & 59121.74 &                               0.00 &                               0.00 &                   29.43, +3.36 &                        (3.49, 38.2, 3.32) &                              0.86 \\
134577:31638233 &  265.17, +5.34 & 59129.92 &                               0.00 &                               0.00 &                             -- &                            (0.00, --, --) &                              0.00 \\
134599:66310113 & 221.22, +14.44 & 59136.09 &                               0.00 &                               0.00 &                 220.96, +14.48 &                         (2.71, 4.3, 1.78) &                              0.75 \\
134621:31008065 & 260.82, +14.55 & 59143.28 &                               0.00 &                               1.09 &                 260.57, +15.11 &                        (4.00, 25.8, 2.81) &                              0.69 \\
134698:40735501 &  105.73, +5.87 & 59167.63 &                               0.00 &                               0.00 &                  105.40, +6.37 &                        (0.70, 14.1, 2.85) &                              0.12 \\
134699:70289682 &  195.12, +1.38 & 59168.09 &                               0.00 &                               1.20 &                  194.62, +2.03 &                        (2.77, 15.2, 2.42) &                              0.40 \\
134715:65785778 & 307.66, +40.72 & 59173.41 &                               0.00 &                               0.55 &                 307.90, +44.55 &                       (14.27, 65.6, 3.99) &                              1.99 \\
134751:31476488 &  30.54, -12.10 & 59183.85 &                               0.00 &                               0.90 &                  31.03, -11.97 &                        (2.14, 12.0, 4.01) &                              0.34 \\
 134777:8912764 &    6.86, -9.25 & 59192.43 &                               0.00 &                               0.00 &                    7.22, -9.50 &                        (8.07, 22.9, 4.16) &                              1.49 \\
134817:29175858 & 261.69, +41.81 & 59204.53 &                               0.00 &                               0.00 &                 261.14, +41.33 &                        (0.56, 14.4, 3.16) &                              0.15 \\
134818:73718836 & 206.37, +13.44 & 59205.04 &                               0.00 &                               0.00 &                 206.19, +13.55 &                         (0.06, 2.9, 2.90) &                              0.19 \\
\enddata
\end{longdeluxetable}

% \begin{acknowledgments}
% \end{acknowledgments}

\bibliography{references}{}
\bibliographystyle{aasjournal}

\end{document}